# Improved Predictive Models for Acute Kidney Injury with IDEAs: Intraoperative Data Embedded Analytics

Lasith Adhikari, PhD[1,4], Tezcan Ozrazgat-Baslanti, PhD[1,4], Paul Thottakkara, MS[4], Ashkan Ebadi, PhD[1,4], Amir Motaei, PhD[1,4], Parisa Rashidi, PhD[2,4], Xiaolin Li, PhD[3,4], Azra Bihorac, MD, MS[1,4]

[1] Division of Nephrology, Hypertension and Renal Transplantation, Department of Medicine, University of Florida, Gainesville, FL, USA

[2] Biomedical Engineering Department, University of Florida, Gainesville, FL, USA

[3] Department of Electrical and Computer Engineering, University of Florida, Gainesville, FL, USA

[4] Precision and Intelligent Systems in Medicine (Prisma[P]), University of Florida, Gainesville, FL, USA

**Corresponding author:** Azra Bihorac MD MS, Department of Medicine, Precision and Intelligent Systems in Medicine (Prisma[P]), Division of Nephrology, Hypertension, and Renal Transplantation, PO Box 100224, Gainesville, FL 32610-0254. Telephone: (352) 294-8580; Fax: (352) 392-5465; Email: abihorac@ufl.edu

**Funding sources:** This work was supported by the National Institute of General Medical Sciences [grant numbers R01 GM110240, P50 GM-111152]; the Clinical and Translational Science Institute, University of Florida [grant number 97071]; the Gatorade grant 127900 from University of Florida; and the NIH/NCATS Clinical and Translational Sciences Award [UL1 TR000064].

**ABSTRACT**

**Purpose:** Acute kidney injury (AKI) − an abrupt loss of kidney function − is a common and serious complication after a surgery which is associated with a high incidence of morbidity and mortality. The majority of existing perioperative AKI risk score prediction models are limited in their generalizability and do not fully utilize the physiological intraoperative time-series data. Thus, there is a need for intelligent, accurate, and robust systems, able to leverage information from large-scale data to predict patient's risk of developing postoperative AKI.

**Materials and Methods:** A retrospective single-center cohort of 2,911 adult patients (age ≥ 18) who underwent surgery at the University of Florida Health from 2000 to 2010 and had a length of stay > 24 hours, has been used for this study. We study the incorporation of intraoperative time-series features to the preoperative prediction risk score can enrich the performance of the predictive models in classifying patients who will develop postoperative AKI. We used machine learning and statistical analysis techniques to develop perioperative models to predict the risk of acute kidney injury (postoperative risk during the first 3 days, 7 days, and until the discharge day) before and after the surgery. In particular, we examined the improvement in risk prediction by incorporating intraoperative physiologic time series data, such as, mean arterial blood pressure (MAP), systolic blood pressure, diastolic blood pressure, minimum alveolar concentration (MAC), and heart rate (HR), etc. For an individual patient, the preoperative model produces a probabilistic AKI risk score, which will be enriched by integrating intraoperative statistical features through a machine learning stacking approach inside a random forest classifier. Using the Yoden's index based cut-off optimization, the model stratifies patients into low and high risk categories. We compared the performance of our model based on the area under the receiver operating characteristics curve (AUROC), accuracy and net reclassification improvement (NRI) metrics.

**Results:** The prevalence of 7day postoperative AKI (i.e., AKI-7day) among 2,911 adult surgical patients was approximately 40% (prevalence of 3day and overall AKI were 34% and 46%, respectively). The predictive performance of the proposed model is better than the preoperative data only model. For AKI-7day outcome: AUC was 0.86 (accuracy was 0.78) from the proposed model, while the preoperative AUC was 0.84 (accuracy 0.76). Furthermore, with the integration of intraoperative features, we were able to classify patients who were misclassified in the preoperative model. According to the NRI statistic, there was a significant improvement in reclassification: NRI for AKI-3day was 8%, AKI-7day was 7%, and AKI-overall was 4%.

**Conclusions:** We proposed an intelligent machine learning model that is able to improve patients' postoperative AKI risk score by taking the intraoperative features into account. Further research can address other post-surgical complications as well as validation of proposed system on external datasets.

**INTRODUCTION**

Acute kidney injury (AKI), previously known as acute renal failure, is one of the most common postoperative complication of many inpatient procedures (1). AKI is associated with increased risk of morbidity, mortality and with high financial costs due to prolonged hospital stay (2, 3). AKI has been growing significantly at a rate of 14% per year since 2001 and the in-hospital deaths due to AKI rose by 16% in the US between 2001 and 2011 (4). Early detection of perioperative AKI risk is clinically challenging and accurate prediction of AKI has garnered a significant attention recently.

A number biomarkers − NGAL, Cys-C, KIM-1, IL-18, L-FABP, etc. (5) − have been proposed for early detection of AKI based on serum, plasma or urine. However, some biomarkers (e.g., NGAL) reflects the severity of disease than being specific to the kidney injury and some of them are not significantly better than the standard clinical evaluations in early stages (6). Also, applying biomarkers to low risk patients will increase the health care cost, hence they are rarely used in everyday practice. With the advancement of technology and availability of abundant electronic health records (EHR), a number of predictive models also have been developed to estimate postoperative AKI risk in different clinical settings (7). Most of the existing AKI risk score calculators are limited to the preoperative factors (8), applicable only to a specific surgery type (9, 10). Some of the available online prognostic calculators (11) are designed only for ICU patients (under surgical or medical category) without taking any surgical features into account. However, several studies have investigated the association between intraoperative data (such as the duration of MAP (Figure 1) in (12) and combination of low haemoglobin and severe hypotension in (13)) and the risk of

AKI. The AKI prediction model in (14) integrates only few intraoperative variables (i.e., procedure duration, fluid balance, plasma and platelet transfusion) and specific to vascular surgeries.

Therefore, the majority of existing perioperative AKI risk models do not fully utilize the available rich physiologic intraoperative data found in EHR. Thus, there is a need for intelligent, accurate, and robust systems, able to leverage information from operative period to predict postoperative AKI risk. The aim of this study was to develop a machine learning algorithm that could integrate the intraoperative features and improve the classification performance of preoperative prediction models.

## METHODS

The method of this study has been designed to evaluate the effectiveness and efficiency of AKI prediction before and after a major surgery with the integration of intraoperative data. The study was designed and approved by the Institutional Review Board of the University of Florida and the University of Florida Privacy Office. The statistical analysis and machine learning were performed using Python, R, and SAS software.

### Data source

Using the University of Florida Integrated Data Repository, we have previously assembled a single center cohort of perioperative patients by integrating multiple existing clinical and administrative databases at UF Health (8). The billing database for UF Health, established in 1990, provides detailed information on patient demographics, outcomes, comprehensive hospital charges, hospital characteristics, insurance status and physician identity. International Classification of Diseases, Ninth Revision, Clinical Modification (ICD-9-CM) codes for up to fifty diagnoses and procedures are listed for each admission. We included all patients admitted to the hospital for longer than 24 hours following any type of operative procedure between January 1, 2000 and November 30, 2010. This dataset was integrated with the laboratory, pharmacy and blood bank databases and intraoperative database (Centricity Perioperative Management and Anesthesia, General Electric Healthcare, Inc.) to create a comprehensive perioperative database for this study.

**Participants**

We identified patients with age greater or equal to 18 years admitted to the hospital for longer than 24 hours following any type of inpatient operative procedure between January 1, 2000 and November 30, 2010. We chose only the first procedure of patients with multiple surgeries for further analysis. We excluded patients with chronic kidney disease (CKD) stage five on admission as identified by the previously validated ICD-9-CM diagnostic and procedure codes and those with missing serum creatinine, resulting in a 2,911 patient population for the analysis. We obtained institutional review board approval through the UF Gainesville Health Science Center Institutional Review Board and UF Privacy Office (#5-2009).

**Outcomes**

Main outcome of interest is postoperative acute kidney injury defined using the recent KDIGO (Kidney Disease: Improving Global Outcomes) criteria. A non end-stage renal disease patient (non-ESRD) will be diagnosed by AKI if one of the following holds: (1) If patient is undergoing a renal replacement therapy; (2) if a measured creatinine value in blood exceeds 1.5 times of the baseline of the creatinine (in blood) for the patient; (3) if the creatinine (in blood) has an increment of at least 0.3mg/dl with in 48-hour period. The baseline creatinine value determines the normal creatinine value for the patient based on the creatinine values measured in the past one year. If the patient is not diagnosed with chronic kidney disease, the modification of diet in renal disease study (MDRD) equation was used determine the baseline creatinine value. The AKI status (yes/no) during the first 3 postoperative days (i.e., AKI-3day), during the 7 postoperative days (i.e., AKI-7day), and during the whole postoperative time up to the discharge date (i.e., AKI-overall) were used as outcome variables in this study.

**Predictor Features**

We derived features for preoperative and intraoperative variables (SDC Table 1) separately. For the preoperative stage, we derived predictor features from available demographic, socio-economic, administrative, clinical, pharmacy and laboratory variables. Preoperative comorbidities were derived using the ICD-9-CM codes as binary variables and with the Charlson comorbidity index. Also, the primary procedure type was modeled based on the ICD-9-CM codes with a forest structure, where each node represents a group of procedures, with roots representing most general groups of procedures and leaf nodes representing specific procedures.

For the intraoperative stage, we derived statistical features such as minimum, mean, maximum, short and long term variability (15) from major physiologic time series

(e.g., mean arterial blood pressure (MAP), systolic blood pressure, diastolic blood pressure, minimum alveolar concentration (MAC), and heart rate (HR)). In addition, abnormal value percentages, value counts, and variances have been used as features for laboratory variables.  Intraoperative medications and other operative characteristics (e.g., procedure durations, anesthesia type, etc.) also have been included to the intraoperative predictors.

**Predictive Analytics Workflow**

The proposed intraoperative data embedded preoperative model consists of two main layers: *Preoperative* and *intraoperative* (Figure 2). Each layer contains three main cores: *Data transformer*, *data engineering*, and *data analytics*.

**Data Transformer:**

Data from various sources are integrated into longitudinal cohorts inside the data transformer layer before direct them to the data engineering layer.

**Data Engineering:**

Primarily, variable generation and data preprocessing were performed in the data engineering layer. After generating new variables from the raw data, we cleaned variables following common set of rules. In particular, for the time series variables in the intraoperative layer, observations were first truncated to fit the corresponding surgery start and stop times for each patient. All extreme values were replaced by average of their five nearest neighbors. Isolated peaks and valleys due to any mechanical malfunctions were replaced by moving average values from the baseline time series. After the cleanup, we extracted various statistical features (e.g., minimum, mean, maximum, short and long term variability) from time series data (15).

Medications and laboratory results during the surgery were also extracted and cleaned for the intraoperative model building. In particular, the outlier detection and removal for all continuous variables were performed by replacing the top and bottom 1% of data using random uniform values generated from 95%-99.5% and 0.5%-5% percentiles, respectively. Categorical variables with more than five levels were modeled with conditional probabilities for a patient to have a particular variable value conditioning on the outcome.

**Data Analytics:**

*Preoperative data only model:* Based on our previous study (16), multivariable modeling of the association between preoperative variables and AKI was performed using generalized additive models (GAM) with logistic link function (17, 18). All models were adjusted for non-linearity of covariates using nonlinear risk functions estimated with thin plate regression splines (19). The best GAM model was picked using a 5-fold cross validation technique and the preoperative prediction scores were generated as the output.

*Intraoperative data embedded preoperative model:* To enrich the predictions from GAM model, here we proposed to use preoperative prediction scores as a new feature to the intraoperative model (stacking technique). All the intraoperative statistical features along with the preoperative prediction scores underwent a univariate analysis and only statistically significant (based on the F-test statistic) features were considered for the random forest (RF) model (20). In particular, the feature selection and other hyper parameters in scikit-learn (21) random forest classifier (i.e., number of trees, maximum features for the best split, minimum number of samples required to be at a leaf node) were tuned simultaneously using a grid search technique with 5 fold cross validation.

**Model Validation**

Along with the preoperative data only model and the intraoperative data incorporated preoperative prediction score model, we trained two other models, i.e., intraoperative data only RF model and full preoperative data embedded intraoperative RF model using 70% of data to evaluate the efficiency and effectiveness of our proposed method.

All the models are validated using the remaining 30% of testing data cohort of size 873 patients. We assessed each model's discrimination using the area under receiver operating characteristic curves (AUROC). For stratification, the optimal cut-off points were calculated based on the maximum Youden's index value (22) computed during the training process for each outcome. As a result, patients were classified to two different categories: low risk and high risk patients. Also, we built the classification table from which we calculated accuracy, sensitivity, specificity, and positive and negative predictive values for each model. In addition, the net reclassification improvement (NRI) index (23) was used to quantify how well our proposed model reclassifies AKI patients over the existing preoperative data only models. We used bootstrap sampling to obtain 95% confidence intervals for all the performance metrics.

**RESULTS**

All input variables are described in SDC Table 1 and overall cohort is summarized based on the outcomes in Table 1 (AKI-7day) and SDC Table 2 (AKI-3day and AKI-overall).

**Preoperative and intraoperative baseline characteristics of participants:**

Among 2,911 patients, approximately 60% of the population were male. The median age of patients who developed AKI-7day was 63 years, in which 65% of them are male. Other than the demographic features, socio-economic characteristic of AKI patients were similar to those with no AKI except the distance from residency to hospital. The prevalence of 7day postoperative AKI was approximately 40% (prevalence of 3day and overall AKI are 34% and 46%, respectively). The distribution of AKI outcomes did not significantly differ between training and testing cohorts: 34% (training) and 35% (testing) for AKI-3day patients; 39% (training) and 41% (testing) for AKI-7day patients; 45% (training) and 48% (testing) for AKI-overall patients. Patients with any degree of presenting comorbidity (from CCI) and especially those with documented CKD (on admission) were more likely to develop AKI -- 80% of CKD patients got AKI during the first 7 postoperative days. Admission features such as weekend admission, admission source, admitting type, and operative features such as surgery type, time to surgery from the admission, total surgery time, and night/day surgery are statistically different (p-value < 0.05) among AKI and no AKI patients. Admission day medications (only except the betablockers) as well as the intraoperative medications (i.e., diuretic and pressors) have a direct impact on getting AKI or no AKI.

According to the physiologic intraoperative time series analysis, patients with AKI-7day were more likely to have a lower base signal mean value for MAP compared to those without AKI. However, the same AKI-7day patients were more likely to have a higher base signal mean value for HR compared to those without AKI. Furthermore, 95% confidence interval for HR values is narrower for patients with AKI (Figure 3).

The predictive performance of the proposed model (i.e., only integrating the preoperative prediction scores with intraoperative features) is better than the existing preoperative data only model (Figure 4A). In particular, for AKI-7day outcome, AUROC is 0.86 for the proposed model, while the preoperative AUROC is 0.84 (accuracy is 0.78 for the proposed model, while the preoperative accuracy is 0.76). The complete performance of the proposed model over the preoperative data only model for the testing cohort with 95% confidence intervals is given in Table 2. The AUROC value comparison for all 4 models (i.e., intraoperative data only model, preoperative data only model, the model with preoperative prediction scores integrated with intraoperative features, and model with the full set of preoperative features integrated with intraoperative features) are given in Figure 5. Even though the AUROC of the full set of preoperative feature integrated model is slightly higher than the proposed model, both models are comparable – training is computationally intense and time consuming for the full set of preoperative feature integrated model.

**Reclassification of risk groups and their characteristics:**

To evaluate the effectiveness of the intraoperative data integrated model, we reclassified the risk groups predicted by preoperative data only model over the proposed model (Figure 4 B-C and SDC Figure 2). The net reclassification improvement (NRI) for AKI-7day between the proposed model and the preoperative data only model is 7% (NRI for AKI-3day is 8% and for AKI-overall is 4%) (Table 2).

Furthermore, patients who have been classified as AKI by the proposed model for the misclassified patients from the preoperative data only model, have similar MAP mean value as in the correctly classified patients (Figure 6 A-B). Similarly, misclassified high risk patients in the preoperative model have similar behavior in the MAP and blood products with the correctly classified patients (Figure 6 C-D).

**DISCUSSION**

In this retrospective single cohort study of 2,911 surgical patients from University of Florida Health between 2000 and 2010, we developed machine learning models and evaluated the effectiveness and efficiency of predicting the most common and serious postoperative complication – acute kidney injury – risk with the integration of intraoperative data with the preoperative data. In particular, here we have developed four different models (i.e., intraoperative data only model, preoperative data only model, intraoperative data integrated preoperative prediction score model, and model with all intraoperative and preoperative features) to measure the performance with each of the three AKI outcomes: AKI-3day, AKI-7day, and AKI-Overall.

This study has more strengths than the existing AKI prediction models. During the surgery, capturing dynamic changes of surgical complications is crucial in providing high quality postoperative care. More personalized preventive management requires timely and accurate synthesis of large amount of clinical data – EHR, vital signs, laboratory results, medications, etc. – throughout the perioperative stages. Specifically, the existing models (9-11) do not fully utilize the available time-varying physiologic data during the surgery. However, the incorporation of statistically crafted intraoperative features through the proposed method will increase the accuracy of predicting AKI by 3% to 8%. According to the feature importance from the fitted random forest model, lactic acid, mean arterial blood pressure, systolic blood pressure, diastolic blood pressure, red cell distribution, platelet count, blood products, heart rate, and surgery time are with in the first 20 most important intraoperative predictors. More importantly, the proposed method captured high AKI risk patients while they were classified as low risk patients from the preoperative data only models (there was a 4% to 8% net reclassification improvement). In addition, this proposed model will be embedded to our existing postoperative prediction intelligent platform in future to perform real time.

This study has some limitations as well. First, we chose only the first surgery of patients with multiple surgeries for building the proposed predictive models. However, there is an opportunity to consider all surgery encounters when building the random forest models. Second, we only predicted AKI outcome in three different time points. We can extend the proposed predictive model with other postoperative outcomes – sepsis, respiratory failure, cardiovascular complications, etc. – as well. Moreover, these models have only been validated by the unseen data from the same cohort. Therefore, further validation is required using an external dataset.

**CONCLUSIONS**

We proposed a machine learning model based on random forest classifier that is able to improve patients' postoperative AKI risk score by taking the intraoperative features into account. Further research can address other post-surgical complications as well as validation of proposed system on external datasets.

**ACKNOWLEDGMENT**

We would like to acknowledge our research colleagues Shivam Mittal, MS, Swati Sisodia, MS, and R. W. M. A. Madushani, PhD with data cleaning and coding, George Omalay for project management.

**Figures and Figure Legends**

Figure 1. Mean arterial blood pressure over the time and their effect in predicting AKI: (A) AKI-3day, (B) AKI-7day, and (C) AKI-Overall. Note that short durations of MAP less than 55 mmHg are associated with AKI.

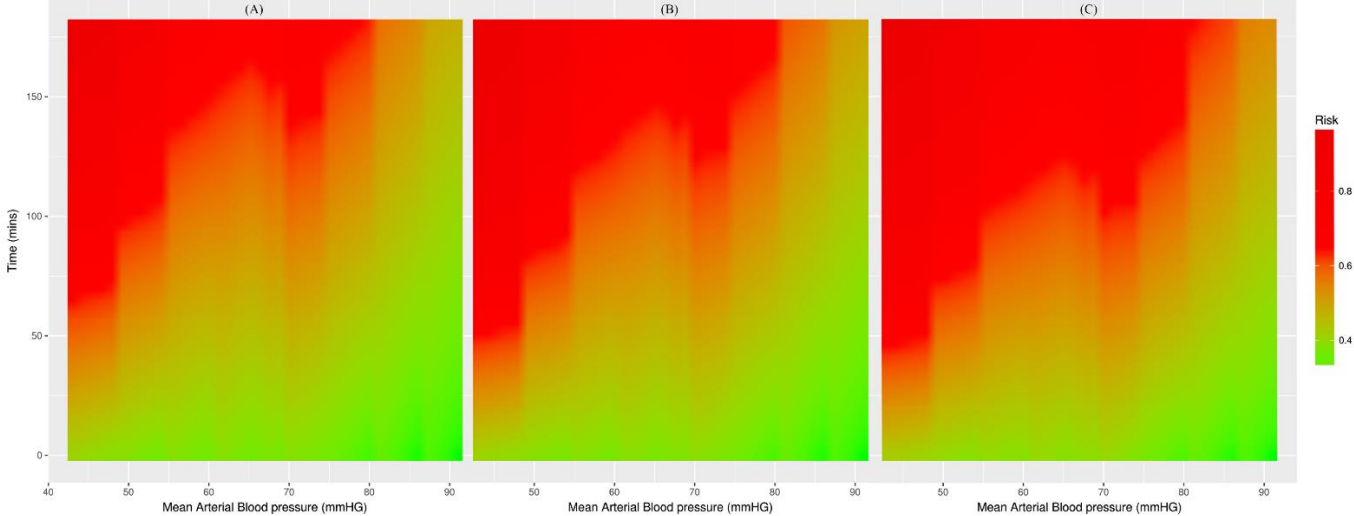

Figure 2. The conceptual diagram of the intraoperative data integrated AKI prediction model. This diagram shows the aggregation of data transformer, data engineering, and data analytics modules in preoperative and intraoperative stages. In particular, we proposed to stack preoperative prediction scores with the cleaned and feature engineered intraoperative data to obtain improved prediction results for acute kidney injury.

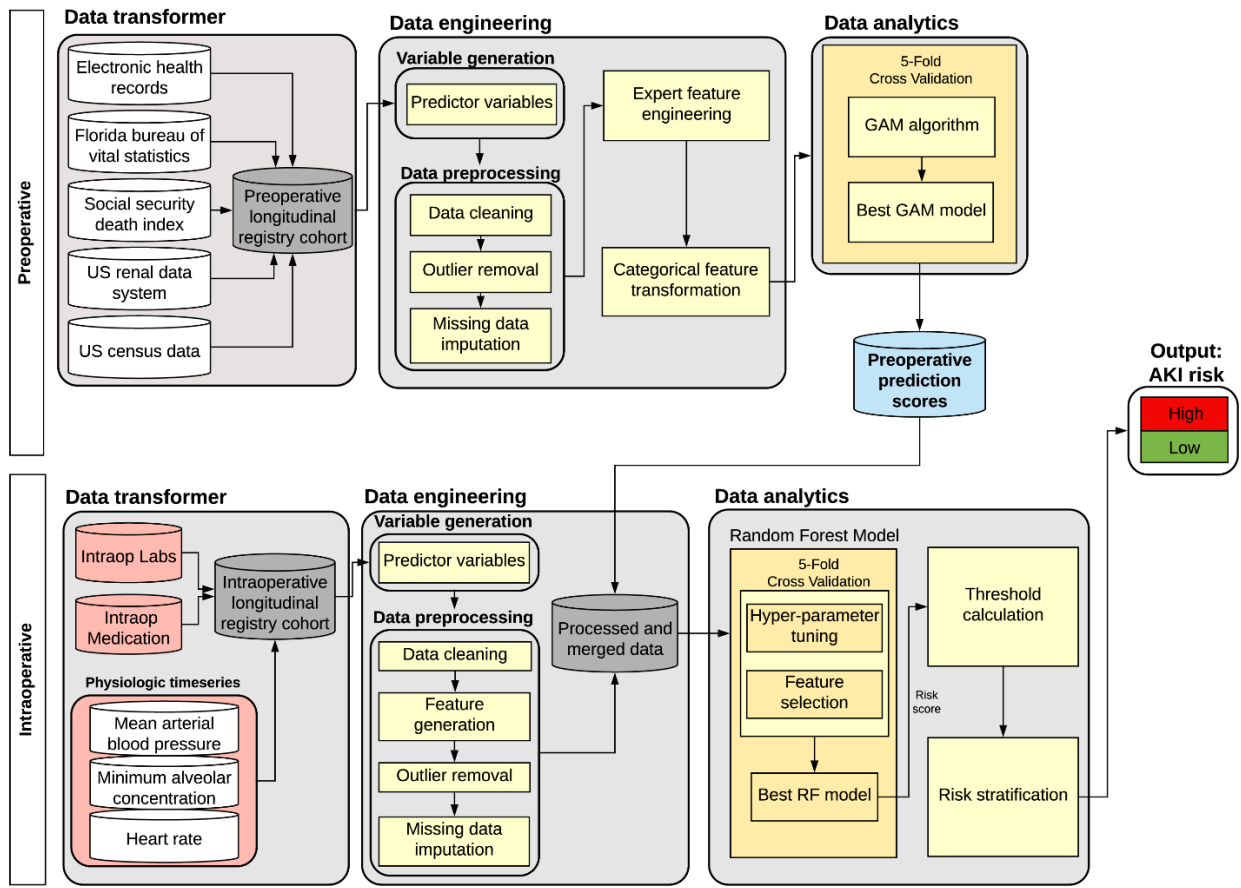

Figure 3. Physiologic time series variations stratified by AKI-7day outcome for 100 randomly selected patients during the first 200min of the surgery. (A) Average of the mean arterial blood pressure including 95% CI for AKI and no AKI, (B) average of the heart rate values including 95% CI for AKI and no AKI, (C) average of the MAC values including 95% CI for AKI and no AKI.

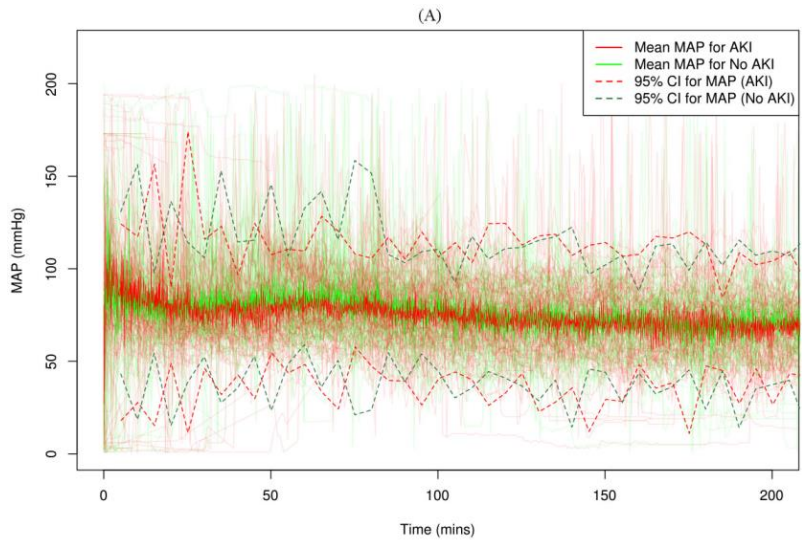
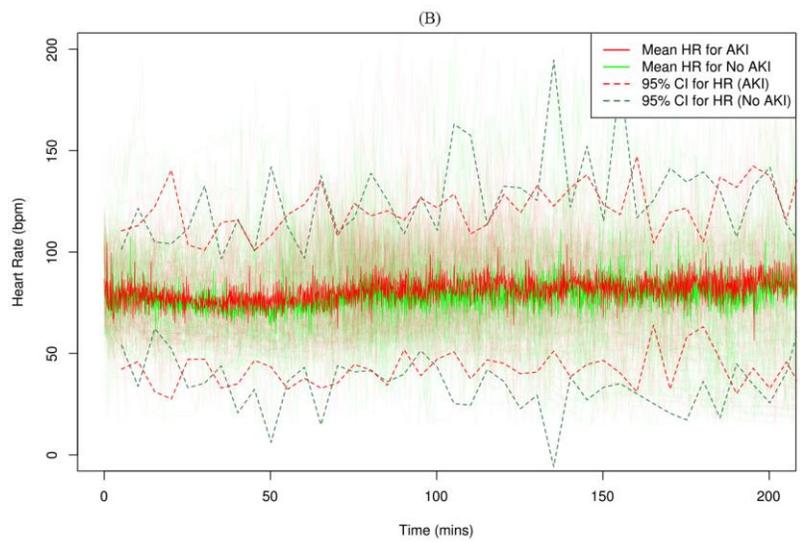
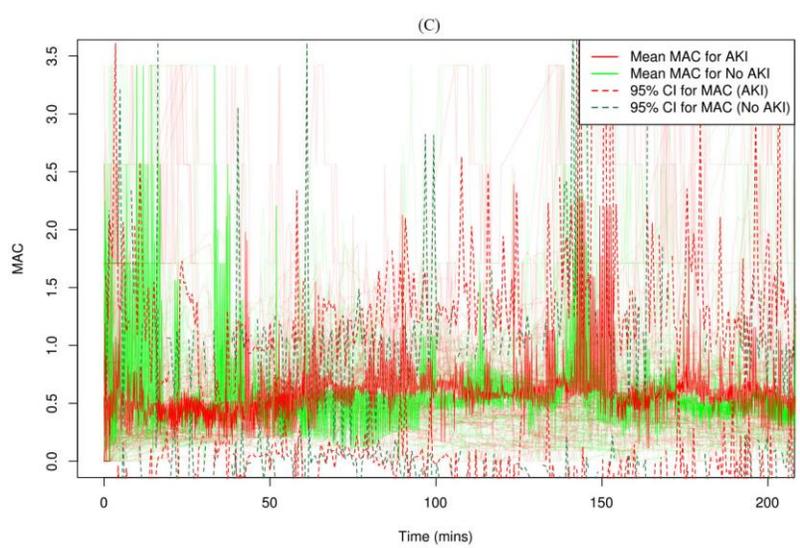

Figure 4. (A) ROC curves for preoperative model (AUC=0.84) and the proposed intraoperative data integrated model (AUC=0.86) with the AKI-7day outcome from the testing cohort. (B) The reclassification of true AKI-7day patients among preoperative model and proposed model. The proposed method reclassified false negative patients from preoperative model as high AKI risk patients. (C) The reclassification of true no AKI-7day patients between preoperative model and proposed model.

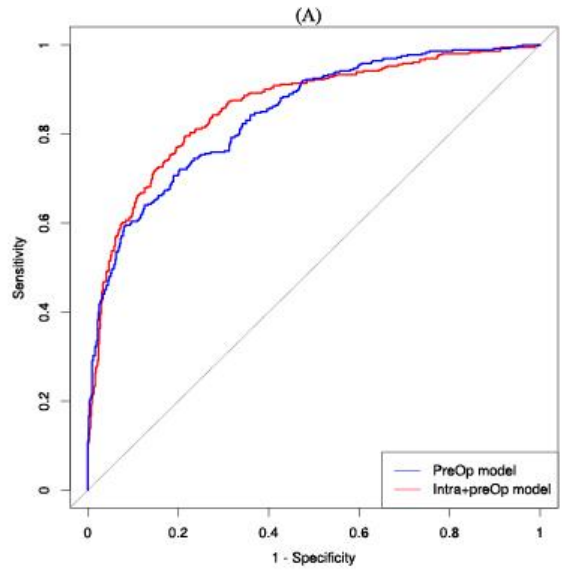
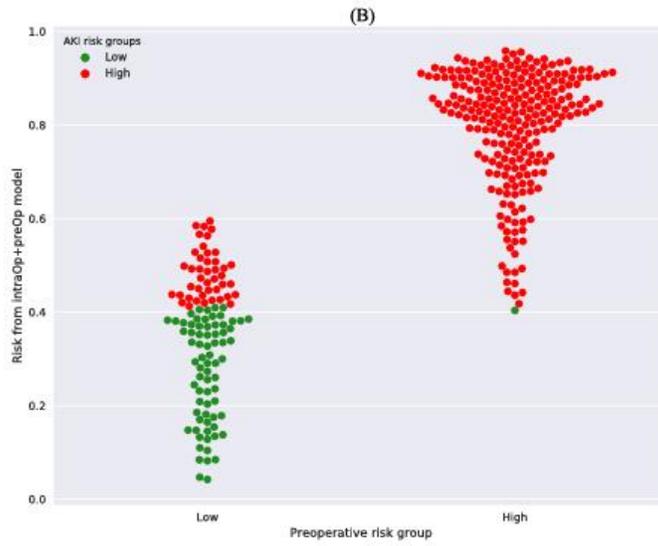
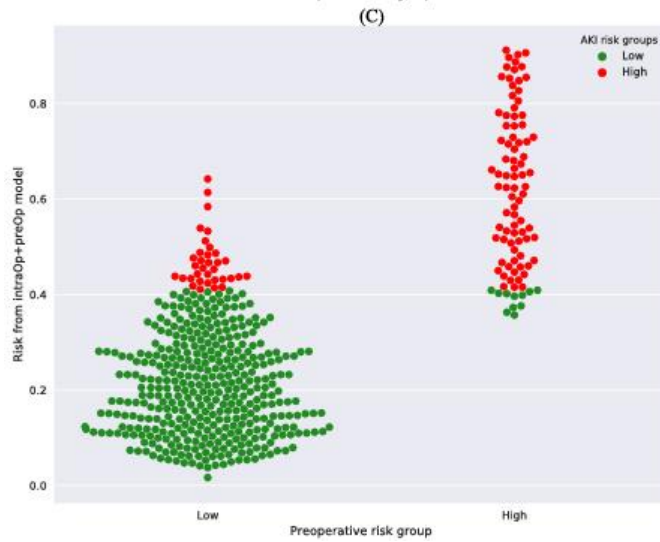

Figure 5. AUROC value comparison among all 4 models: Intraoperative data only model, preoperative data only model, intraoperative data integrated preoperative prediction score model (proposed model), and model with all intraoperative and preoperative data for all three AKI outcomes: AKI-3day, AKI-7day, and AKI-Overall from the testing cohort.

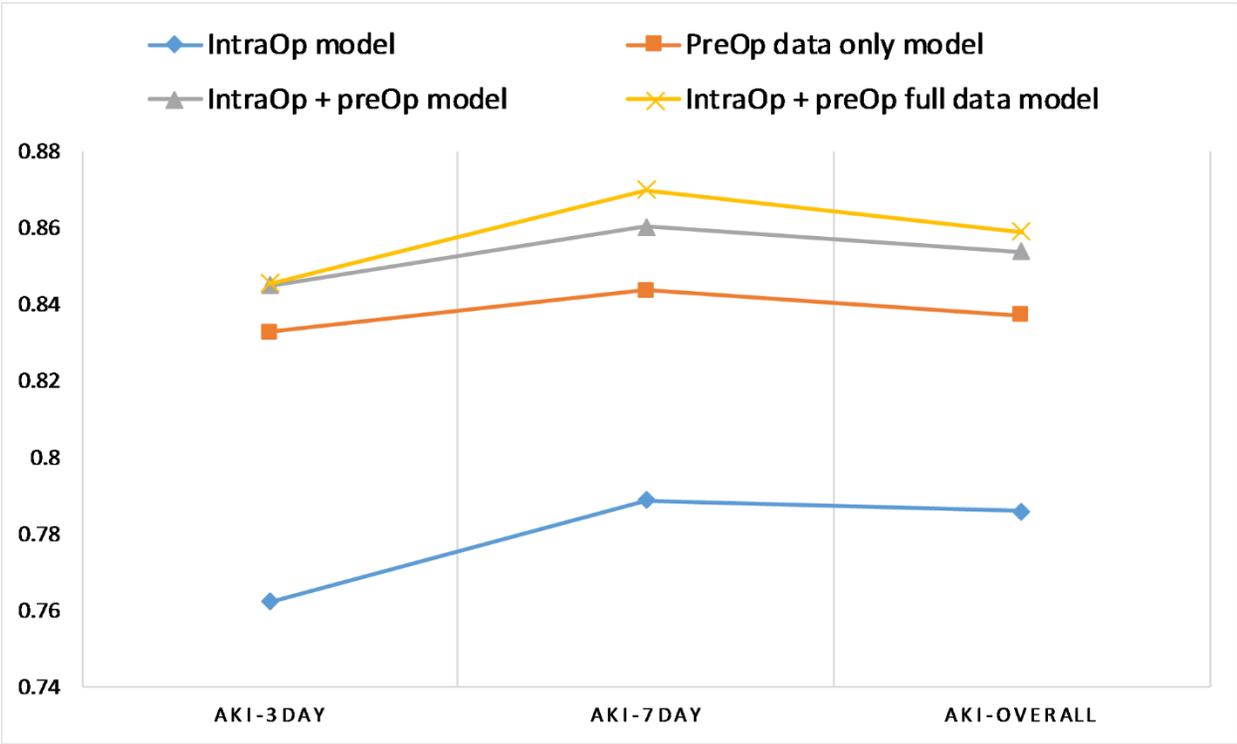

Figure 6. True AKI patient characteristics after reclassification. (A) Reclassification from proposed model for misclassified patients in preoperative model. (B) Reclassification from proposed model for correctly classified patients in preoperative model. Mean MAP of high risk patients in (A) is statistically similar to mean MAP in (B) (p-value=0.75 > 0.05 by the Wilcoxon rank sum test). (C) Blood products (ml) vs. mean MAP for misclassified patients in preoperative model. (D) Blood products (ml) vs. mean MAP for correctly classified patients in preoperative model. Mean values of blood products and mean MAP for high risk patients in proposed model are statistically similar between (C) and (D). Note: Box represents 95% confidence intervals of the mean for the given variable.

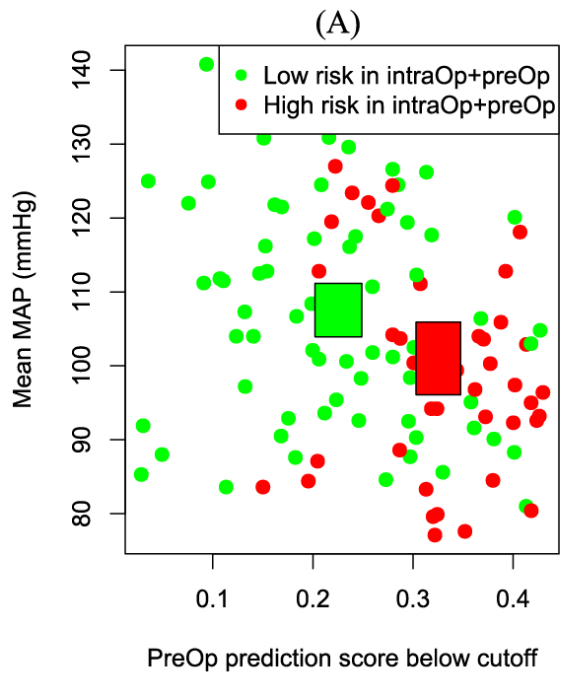 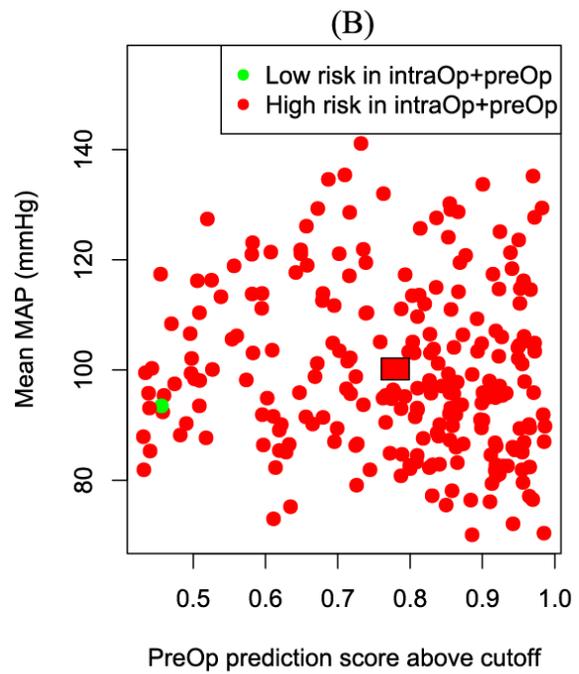 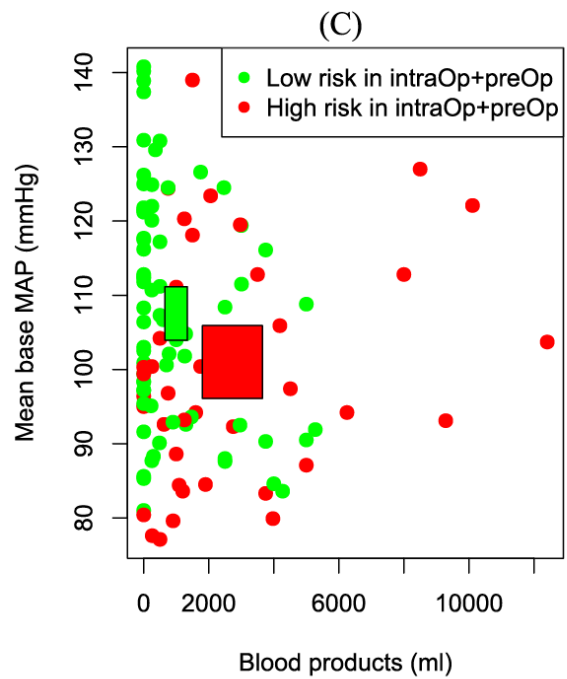 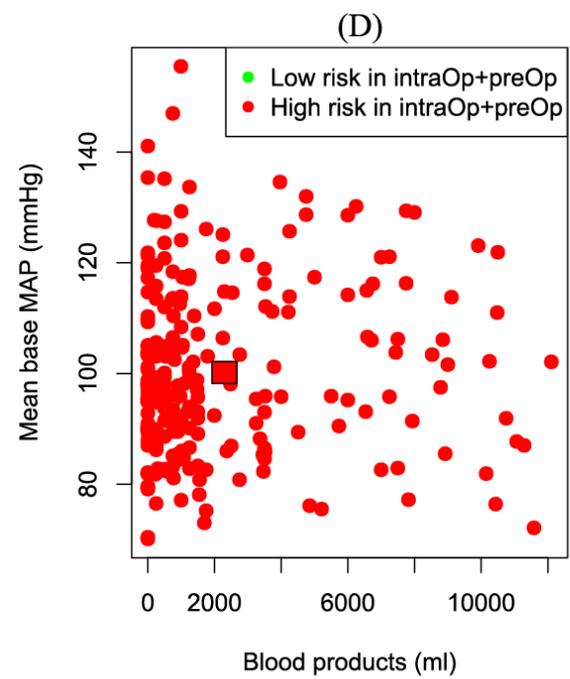

## Tables

Table 1. Summary of overall cohort stratified by acute kidney injury - 7day.

|  | Overall (N=2,911) | Acute Kidney Injury - 7day | |
|---|---|---|---|
|  |  | Yes (N=1,163) | No (N=1,748) |
| **Demographic features** | | | |
| Age, median (25th-75th) | 60.0 (49.0,69.0) | 63.0 (52.0,72.0)[a] | 58.0 (47.0,67.0) |
| Male gender, n(%) | 1760 (60.46%) | 753 (64.75%)[a] | 1007 (57.61%) |
| Race, n(%) | | | |
|   White | 2374 (81.55%) | 934 (80.31%) | 1440 (82.38%) |
|   AA | 265 (9.1%) | 115 (9.89%) | 150 (8.58%) |
|   Hispanic | 117 (4.02%) | 53 (4.56%) | 64 (3.66%) |
|   Missing | 87 (2.99%) | 37 (3.18%) | 50 (2.86%) |
|   Other | 68 (2.34%) | 24 (2.06%) | 44 (2.52%) |
| Primary insurance, n(%)[a] | | | |
|   Private | 1208 (41.5%) | 411 (35.34%) | 797 (45.59%) |
|   Medicare | 1204 (41.36%) | 569 (48.93%) | 635 (36.33%) |
|   Medicaid | 340 (11.68%) | 142 (12.21%) | 198 (11.33%) |
|   Uninsured | 159 (5.46%) | 41 (3.53%) | 118 (6.75%) |
| Zip (top 3 codes), n(%)[a] | | | |
|   32608 | 34 (1.17%) | 12 (1.03%) | 22 (1.26%) |
|   32060 | 27 (0.93%) | 12 (1.03%) | 15 (0.86%) |
|   32605 | 27 (0.93%) | 6 (0.52%) | 21 (1.2%) |
| **Socio-economic features** | | | |
| Neighborhood characteristics | | | |
| Rural area, n(%) | 767 (26.35%) | 300 (25.8%) | 467 (26.72%) |
| Total population, median (25th-75th) | 19162 (10639,30611) | 19287 (11056,30533) | 18931 (10510,30611) |
| Median income, median (25th-75th) | 34372 (29980,41410) | 34459 (30084,41410) | 34328 (29854,41410) |
| Total proportion of African-Americans (%), median (25th-75th) | 9.6 (3.9,17.6) | 9.6(3.7,19.5) | 9.5 (3.9,16.5) |
| Total proportion of Hispanic (%), median (25th-75th) | 4.2 (2.5,6.8) | 4.1 (2.5,7.1) | 4.3 (2.6,6.7) |
| County (top 3 categories), n(%) | | | |
|   Alachua | 383 (13.16%) | 139 (11.95%) | 244 (13.96%) |
|   Marion | 250 (8.59%) | 73 (6.28%) | 177 (10.13%) |
|   Georgia | 142 (4.88%) | 63 (5.42%) | 79 (4.52 %) |
| Distance from residency to hospital (km), median (25th-75th) | 68.01 (28.9,143.19) | 72.73 (30.69,153.35)[a] | 61.02 (28.2,131.48) |
| Population proportion below poverty (%), median (25th-75th) | 12.0 (8.1,17.3) | 11.8 (8.0,17.4) | 12.1 (8.3,17.2) |

**Preoperative characteristics of patients stratified by acute kidney injury**

**Comorbidity features**

| | | | |
|---|---|---|---|
| Charlson's comorbidity index (CCI), median (25th-75th) | 2.0 (1.0,3.0) | 2.0 (1.0,3.0)[a] | 1.0 (0.0,3.0) |
| Chronic kidney disease, n(%) | 346 (11.89%) | 273 (23.47%)[a] | 73 (4.18%) |

**Operative features**

**Admission**

| | | | |
|---|---|---|---|
| Weekend admission, n(%) | 472 (16.21%) | 217 (18.66%)[a] | 255 (14.59%) |
| Admission source, n(%)[a] | | | |
|   Outpatient setting | 1753 (60.53%) | 648 (55.81%) | 1105 (63.69%) |
|   Emergency room | 583 (20.13%) | 221 (19.04%) | 362 (20.86%) |
|   Transfer | 560 (19.34%) | 292 (25.15%) | 268 (15.45%) |
| Admission month (top 3 categories), n(%) | | | |
|   September | 279 (9.58%) | 111 (9.54%) | 168 (9.61%) |
|   October | 273 (9.38%) | 99 (8.51%) | 174 (9.95%) |
|   June | 254 (8.73%) | 92 (7.91%) | 162 (9.27%) |
| Number of operating surgeons, n | 129 | | |
| Number of procedures per operating Surgeon, n(%)[a] | | | |
|   First rank | 422 (14.5%) | 241 (20.72%) | 181 (10.35%) |
|   Second rank | 267 (9.17%) | 70 (6.02%) | 197 (11.27%) |
|   Third rank | 258 (8.86%) | 129 (11.09%) | 129 (7.38%) |
| Admitting type, n(%)[a] | | | |
|   surgery | 2545 (87.43%) | 957 (82.29%) | 1588 (90.85%) |
|   medicine | 366 (12.57%) | 206 (17.71%) | 160 (9.15%) |
| Emergent surgery status, n (%)[a] | 1352 (46.44%) | 604 (51.93%) | 748 (42.79%) |
| Surgery type, n(%)[a] | | | |
|   Cardiothoracic surgery | 1415 (48.61%) | 739 (63.54%) | 676 (38.67%) |
|   Other surgeries | 826 (28.38%) | 292 (25.11%) | 534 (30.55%) |
|   Neurologic Surgery | 301 (10.34%) | 30 (2.58%) | 271 (15.5%) |
|   Specialty Surgeries | 243 (8.35%) | 56 (4.82%) | 187 (10.7%) |
|   Non-Cardiac General Surgery | 126 (4.33%) | 46 (3.96%) | 80 (4.58%) |
| Time of surgery from admission (days) | 0.0 (0.0,2.0) | 1.0 (0.0,4.0)[a] | 0.0 (0.0,1.0) |

**Admission day medications, n(%)**

| | | | |
|---|---|---|---|
| Diuretics | 600 (20.61%) | 315 (27.09%)[a] | 285 (16.3%) |
| Bicarbonate | 295 (10.13%) | 151 (12.98%)[a] | 144 (8.24%) |
| Angiotensin-Converting-Enzyme Inhibitors | 351 (12.06%) | 169 (14.53%)[a] | 182 (10.41%) |
| Antiemetic | 1413 (48.54%) | 467 (40.15%)[a] | 946 (54.12%) |
| Betablockers | 872 (29.96%) | 359 (30.87%) | 513 (29.35%) |
| Statin | 502 (17.24%) | 230 (19.78%)[a] | 272 (15.56%) |
| Pressors or inotropes | 424 (14.57%) | 223 (19.17%)[a] | 201 (11.5%) |

***Intraoperative characteristics of patients stratified by acute kidney injury***

**Physiologic intraoperative time series variables, median (25th-75th)**

| | | | |
|---|---|---|---|
| Diastolic BP - maximum | 140 (121,140) | 140 (125,140)[a] | 140 (120,140) |
| Diastolic BP - minimum | 20.0 (15.0,25.0) | 20.0 (12.0,20.0)[a] | 20.0 (17.0,30.0) |
| Diastolic BP - average | 58.7 (53.8,65.1) | 56.4 (51.85,62.0)[a] | 60.6 (55.6,67.0) |
| Diastolic BP - long-term variability | 10.4 (8.7,12.7) | 10.7 (9.0,12.9)[a] | 10.2 (8.4,12.6) |
| Diastolic BP - short-term variability | 4.8 (3.8,5.9) | 5.0 (4.0,6.0)[a] | 4.7 (3.7,5.8) |
| Systolic BP - maximum | 240.0 (209.0,240.0) | 240.0 (215.0,240.0)[a] | 240.0 (205.0,240.0) |
| Systolic BP - minimum | 40.0 (31.0,50.0) | 40.0 (28.0,40.0)[a] | 40.0 (34.0,57.0) |
| Systolic BP - average of base signal | 106.1 (93.8,118.65) | 100.1 (90.3,112.9)[a] | 110.9 (97.2,120.72) |
| Systolic BP - long-term variability | 19.6 (15.6,24.2) | 20.8 (16.9,25.1)[a] | 18.6 (15.0,23.3) |
| Systolic BP - short-term variability | 7.7 (6.1,9.5) | 7.9 (6.4,9.6)[a] | 7.7 (5.9,9.4) |
| Mean Arterial BP - maximum | 240 (209.0,240.0) | 240 (215.0,240.0)[a] | 240 (205.0,240.0) |
| Mean Arterial BP - minimum | 40.0 (31.0,50.0) | 40.0 (28.0,40.0)[a] | 40.0 (34.0,57.0) |
| Mean Arterial BP - average | 106.1 (93.8,118.65) | 100.1 (90.3,112.9)[a] | 110.9 (97.2,120.72) |
| Mean Arterial BP - long-term variability | 19.6 (15.6,24.2) | 20.8 (16.9,25.1)[a] | 18.6 (15.0,23.3) |
| Mean Arterial BP - short-term variability | 7.7 (6.1,9.5) | 7.9 (6.4,9.6)[a] | 7.7 (5.9,9.4) |
| Heart rate - maximum | 146 (125.0,171.0) | 148 (128.0,175.0)[a] | 144 (123.95,167.0) |
| Heart rate - minimum | 34.0 (18.0,53.0) | 26.0 (17.0,47.5)[a] | 39.0 (21.0,55.0) |
| Heart rate - average | 82.8 (73.8,92.7) | 84.1 (74.55,94.0)[a] | 81.9 (73.3,91.72) |
| Heart rate - long-term variability | 12.4 (8.6,19.3) | 14.4 (9.35,21.8)[a] | 11.5 (8.5,17.5) |
| Heart rate - short-term variability | 5.5 (4.2,6.6) | 5.8 (4.5,6.8)[a] | 5.3 (3.9,6.5) |
| MAC - maximum | 2.59 (1.71,3.42) | 3.42 (1.97,3.42)[a] | 2.56 (1.71,3.42) |
| MAC - average | 0.6 (0.5,0.7) | 0.5 (0.5,0.7) | 0.6 (0.5,0.7) |
| MAC - long-term variability | 0.3 (0.2,0.4) | 0.3 (0.2,0.5)[a] | 0.3 (0.2,0.4) |

| Intraoperative laboratory results, median (25th-75th) | | | |
|---|---|---|---|
| Carboxyhemoglobin in arterial - maximum | 2.2 (1.4,3.0) | 2.3 (1.45,3.1)[a] | 2.1 (1.3,2.9) |
| Carboxyhemoglobin in arterial - mean | 1.95 (1.3,2.75) | 2.1 (1.4,2.87)[a] | 1.85 (1.2,2.67) |
| Carboxyhemoglobin in arterial - minimum | 1.6 (1.1,2.55) | 1.8 (1.2,2.7)[a] | 1.5 (1.0,2.5) |
| Carboxyhemoglobin in arterial - variance | 0.0 (0.0,0.08) | 0.0 (0.0,0.077) | 0.0 (0.0,0.08) |
| Bicarbonate in Arterial - maximum | 23.3 (21.7,25.0) | 23.3 (21.7,25.2) | 23.3 (21.7,24.9) |
| Bicarbonate in Arterial - mean | 22.65 (21.18,24.4) | 22.65 (21.06,24.4) | 22.65 (21.2,24.4) |
| Bicarbonate in Arterial - minimum | 22.0 (20.4,24.05) | 22.0 (20.25,24.0) | 22.1 (20.6,24.1) |
| Bicarbonate in Arterial - variance | 0.0 (0.0,0.84) | 0.0 (0.0,1.22) | 0.0 (0.0,0.72) |
| Hematocrit - maximum | 31.5 (28.7,34.9) | 31.4 (28.5,34.8) | 31.6 (28.8,35.0) |
| Hematocrit - mean | 30.8 (28.0,34.1) | 30.6 (27.78,33.8)[a] | 30.9 (28.16,34.3) |
| Hematocrit - minimum | 30.2 (27.1,33.6) | 29.9 (26.7,33.3)[a] | 30.5 (27.4,33.9) |
| Hematocrit - variance | 0.0 (0.0,1.12) | 0.0 (0.0,1.28) | 0.0 (0.0,1.12) |
| Hemoglobin in arterial - maximum | 11.0 (10.0,12.2) | 11.0 (9.9,12.2) | 11.1 (10.1,12.2) |
| Hemoglobin in arterial - mean | 10.55 (9.6,11.8) | 10.45 (9.5,11.64)[a] | 10.7 (9.7,11.9) |
| Hemoglobin in arterial - minimum | 10.2 (9.0,11.6) | 10.1 (8.8,11.5)[a] | 10.3 (9.2,11.7) |
| Hemoglobin in arterial - variance | 0.0 (0.0,0.44) | 0.0 (0.0,0.56) | 0.0 (0.0,0.4) |
| Lactic acid - maximum | 2.5 (1.5,4.2) | 3.4 (2.0,5.2)[a] | 2.1 (1.3,3.32) |
| Lactic acid - mean | 2.2 (1.4,3.57) | 2.83 (1.8,4.6)[a] | 1.77 (1.2,2.82) |
| Lactic acid - minimum | 1.6 (1.1,2.9) | 2.2 (1.3,4.0)[a] | 1.4 (1.0,2.4) |
| Lactic acid - variance | 0.0 (0.0,0.25) | 0.0 (0.0,0.47) | 0.0 (0.0,0.18) |
| Mean corpuscular hemoglobin concentration - maximum | 34.5 (33.5,35.5) | 34.5 (33.6,35.5) | 34.4 (33.5,35.5) |
| Mean corpuscular hemoglobin concentration - mean | 34.3 (33.4,35.3) | 34.3 (33.5,35.3) | 34.3 (33.3,35.3) |
| Mean corpuscular hemoglobin concentration - minimum | 34.1 (33.2,35.1) | 34.2 (33.3,35.1) | 34.1 (33.2,35.1) |
| Mean corpuscular hemoglobin - maximum | 30.6 (29.4,31.8) | 30.5 (29.4,31.8) | 30.6 (29.4,31.8) |
| Mean corpuscular hemoglobin - mean | 30.4 (29.26,31.67) | 30.4 (29.3,31.6) | 30.45 (29.25,31.7) |
| Mean corpuscular hemoglobin - minimum | 30.3 (29.1,31.5) | 30.2 (29.1,31.5) | 30.3 (29.1,31.52) |
| Mean corpuscular hemoglobin - variance | 0.0 (0.0,0.04) | 0.0 (0.0,0.04) | 0.0 (0.0,0.05) |
| Mean corpuscular volume - maximum | 88.6 (85.2,92.2) | 88.7 (85.25,92.2) | 88.4 (85.2,92.2) |
| Mean corpuscular volume - mean | 88.23 (85.0,91.8) | 88.3 (85.0,91.59) | 88.2 (85.1,91.94) |
| Mean corpuscular volume - minimum | 87.9 (84.7,91.4) | 87.9 (84.7,91.0) | 87.9 (84.8,91.7) |
| Methemoglobin - max | 0.8 (0.6,1.0) | 0.8 (0.6,1.0)[a] | 0.8 (0.6,0.9) |
| Methemoglobin - mean | 0.7 (0.52,0.9) | 0.7 (0.55,0.9)[a] | 0.7 (0.5,0.8) |
| Methemoglobin - minimum | 0.6 (0.4,0.8) | 0.6 (0.4,0.9)[a] | 0.6 (0.4,0.8) |
| Methemoglobin - variance | 0.0 (0.0,0.023) | 0.0 (0.0,0.024) | 0.0 (0.0,0.022) |
| Mean Platelet Volume - maximum | 8.0 (7.5,8.7) | 8.2 (7.6,8.8)[a] | 7.9 (7.4,8.5) |
| Mean Platelet Volume - mean | 7.9 (7.4,8.5) | 8.07 (7.5,8.69)[a] | 7.8 (7.3,8.4) |
| Mean Platelet Volume - minimum | 7.8 (7.3,8.4) | 7.9 (7.4,8.55)[a] | 7.7 (7.2,8.3) |
| Mean Platelet Volume - variance | 0.0 (0.0,0.03) | 0.0 (0.0,0.02) | 0.0 (0.0,0.04) |
| O2 Content, Arterial - maximum | 15.3 (14.0,16.95) | 15.2 (13.8,16.9)[a] | 15.4 (14.1,17.0) |
| O2 Content, Arterial - mean | 14.7 (13.4,16.3) | 14.5 (13.2,16.1)[a] | 14.8 (13.5,16.4) |
| O2 Content, Arterial - minimum | 14.2 (12.5,16.0) | 14.0 (12.25,15.8)[a] | 14.3 (12.8,16.1) |
| O2 Content, Arterial - variance | 0.0 (0.0,0.91) | 0.0 (0.0,1.03) | 0.0 (0.0,0.81) |
| O2 saturation - max | 99.8 (99.0,100) | 99.9 (99.0,100) | 99.8 (99.0,100) |
| O2 saturation - mean | 99.5 (98.4,100) | 99.55 (98.41,100) | 99.41 (98.4,100) |
| O2 saturation - minimum | 99.3 (97.8,100) | 99.4 (97.8,100) | 99.2 (97.8,100) |
| O2 saturation - variance | 0.0 (0.0,0.02) | 0.0 (0.0,0.01) | 0.0 (0.0,0.02) |
| Partial pressure of carbon dioxide - maximum | 42.8 (38.4,47.2) | 43.1 (38.85,47.5) | 42.8 (38.2,47.0) |
| Partial pressure of carbon dioxide - mean | 40.8 (36.9,45.2) | 40.86 (37.2,45.38)[a] | 40.8 (36.7,45.0) |
| Partial pressure of carbon dioxide - minimum | 38.9 (34.55,44.6) | 39.5 (34.8,44.8)[a] | 38.8 (34.38,44.4) |
| Partial pressure of carbon dioxide - variance | 0.0 (0.0,6.46) | 0.0 (0.0,6.48) | 0.0 (0.0,6.29) |
| PF-ratio | 90.4 (74.87,118.68) | 87.7 (73.87,114)[a] | 91.92 (75.64,122.31) |
| pH - maximum | 7.38 (7.33,7.43) | 7.38 (7.32,7.43) | 7.38 (7.34,7.43) |
| pH - mean | 7.36 (7.32,7.4) | 7.36 (7.32,7.4)[a] | 7.36 (7.33,7.41) |
| pH - minimum | 7.34 (7.3,7.39) | 7.34 (7.3,7.38)[a] | 7.34 (7.31,7.39) |
| Platelet count - maximum | 180 (133,239) | 163 (117,209)[a] | 197 (145,253) |
| Platelet count - mean | 175 (128,229) | 156 (110,203)[a] | 190 (140,245) |
| Platelet count - minimum | 171 (121,223) | 151 (104,198)[a] | 184 (135,240) |
| Platelet count - variance | 0.0 (0.0,78.1) | 0.0 (0.0,60.5)[a] | 0.0 (0.0,84.9) |
| Red blood cells - maximum | 3.59 (3.23,3.97) | 3.57 (3.22,3.94) | 3.6 (3.25,3.98) |
| Red blood cells - mean | 3.5 (3.16,3.87) | 3.46 (3.12,3.85)[a] | 3.52 (3.19,3.9) |
| Red blood cells - minimum | 3.42 (3.07,3.83) | 3.38 (3.0,3.8)[a] | 3.45 (3.11,3.85) |
| Red blood cells - variance | 0.0 (0.0,0.02) | 0.0 (0.0,0.02) | 0.0 (0.0,0.01) |
| Red cell distribution width - maximum | 14.9 (13.9,16.1) | 15.3 (14.4,16.7)[a] | 14.6 (13.7,15.7) |
| Red cell distribution width - mean | 14.8 (13.84,15.95) | 15.2 (14.4,16.5)[a] | 14.5 (13.6,15.5) |
| Red cell distribution width - minimum | 14.7 (13.8,15.9) | 15.1 (14.2,16.4)[a] | 14.4 (13.6,15.5) |
| Red cell distribution width - variance | 0.0 (0.0,0.01) | 0.0 (0.0,0.02) | 0.0 (0.0,0.01) |
| White blood cells - maximum | 13.6 (9.85,18.3) | 14.0 (9.8,19.25)[a] | 13.3 (9.9,17.8) |
| White blood cells - mean | 12.85 (9.3,17.52) | 13.3 (9.2,18.3)[a] | 12.51 (9.33,16.9) |
| White blood cells - min | 12.1 (8.5,16.9) | 12.6 (8.4,17.7)[a] | 11.7 (8.5,16.3) |
| White blood cells - variance | 0.0 (0.0,0.85) | 0.0 (0.0,0.54)[a] | 0.0 (0.0,1.07) |
| Total blood products in ml | 318 (0.0,1250) | 750.0 (0.0,2500)[a] | 0.0 (0.0,750) |
| Total estimated blood loss in ml | 150 (150,500) | 150 (150,350)[a] | 150 (150,500) |
| Total fluids in ml | 2300 (1400,3600) | 2200 (1250,3500)[a] | 2435 (1500,3700) |
| Total urine output in ml | 650 (300,1150) | 700 (350,1200)[a] | 600 (300,1100) |
| **Intraoperative medications, n(%)** | | | |
| Diuretic (vs. No) | 329 (11.3%) | 210 (18.06%)[a] | 119 (6.81%) |
| Pressors (vs. No) | 1942 (66.71%) | 858 (73.77%)[a] | 1084 (62.01%) |
| **Other variables** | | | |
| Total surgery time in mins, median (25th-75th) | 387 (293,483) | 425 (340,521)[a] | 357 (271,449) |
| Night surgery (vs. No), n(%) | 394 (13.53%) | 184 (15.82%)[a] | 210 (12.01%) |
| General anesthesia (vs. BL/BM), n(%) | 2881 (98.97%) | 1153 (99.14%) | 1728 (98.86%) |

[a] p-value < 0.05 compared to no AKI-7day group.

Table 2: Performance of the proposed intraoperative data integrated model over the preoperative data only model for the testing cohort with 95% confidence intervals.

| Outcome | Model | AUC | Accuracy | Sensitivity | Specificity | PPV | NPV | NRI |
|---|---|---|---|---|---|---|---|---|
| AKI-3Day | Preop model | 0.833 [0.804, 0.861] | 0.703 [0.674, 0.732] | 0.824 [0.778, 0.866] | 0.638 [0.600, 0.677] | 0.551 [0.504, 0.596] | 0.870 [0.833, 0.900] | 8.14% [0.04, 0.13] $p<0.05$ |
| | Proposed model | 0.845 [0.817, 0.873] | 0.771 [0.739, 0.796] | 0.775 [0.729, 0.820] | 0.769 [0.734, 0.803] | 0.644 [0.595, 0.693] | 0.863 [0.830, 0.893] | |
| AKI-7Day | Preop model | 0.844 [0.818, 0.870] | 0.762 [0.732, 0.786] | 0.681 [0.637, 0.731] | 0.818 [0.783, 0.849] | 0.726 [0.675, 0.770] | 0.785 [0.745, 0.818] | 7.14% [0.03, 0.12] $p<0.05$ |
| | Proposed model | 0.860 [0.835, 0.886] | 0.784 [0.753, 0.808] | 0.798 [0.756, 0.837] | 0.773 [0.738, 0.809] | 0.713 [0.671, 0.757] | 0.844 [0.809, 0.874] | |
| AKI-Overall | Preop model | 0.837 [0.811, 0.863] | 0.764 [0.734, 0.792] | 0.697 [0.654, 0.742] | 0.826 [0.798, 0.859] | 0.787 [0.742, 0.827] | 0.747 [0.708, 0.785] | 3.82% [0.002, 0.08] $p<0.05$ |
| | Proposed model | 0.854 [0.828, 0.879] | 0.778 [0.748, 0.803] | 0.795 [0.757, 0.833] | 0.762 [0.723, 0.800] | 0.755 [0.714, 0.792] | 0.801 [0.759, 0.836] | |

**Supplemental Digital Content**

SDC Figure 1. The flow diagram of the random forest model used to train and test the perioperative data along with three different AKI outcomes: AKI-3day, AKI-7day, and AKI-Overall. The cohort of size 2,911 with 231 features were randomly split in to 70% train and 30% test cohorts. A random forest classifier was used to train the AKI predication model (we used 5-fold CV for hyper parameter tuning and feature selection) for all three outcomes and performance were tested using the testing cohort.

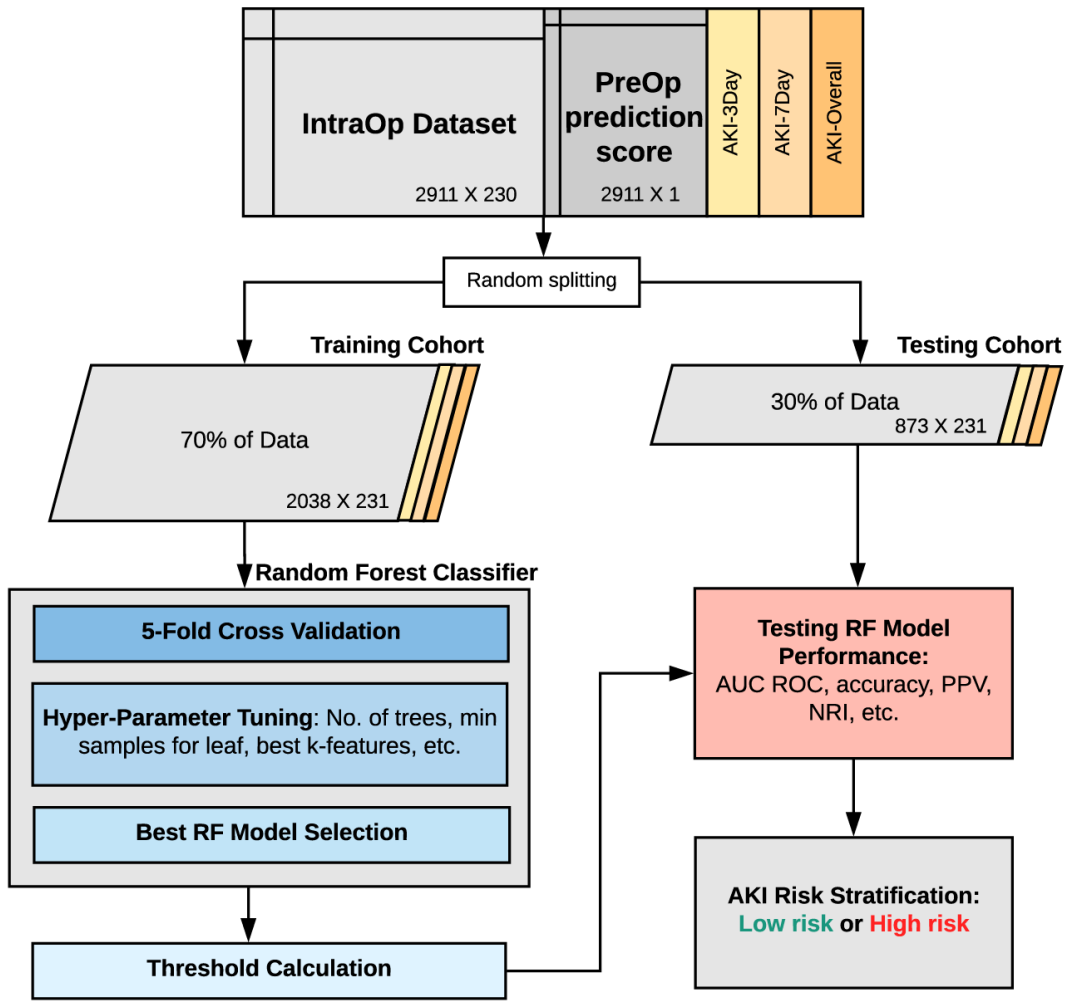

SDC Figure 2. (A) ROC curves for preoperative model (AUC=0.83) and the proposed intraoperative data integrated model (AUC=0.85) with the AKI-3day outcome. (B-C) The reclassification patients among preoperative model and proposed model for AKI-3day and no AKI-3day patients, respectively. (D) ROC curves for preoperative model (AUC=0.84) and the proposed intraoperative data integrated model (AUC=0.85) with the AKI-overall outcome. (E-F) The reclassification of patients between preoperative model and proposed model for AKI-overall and no AKI-overall patients, respectively.

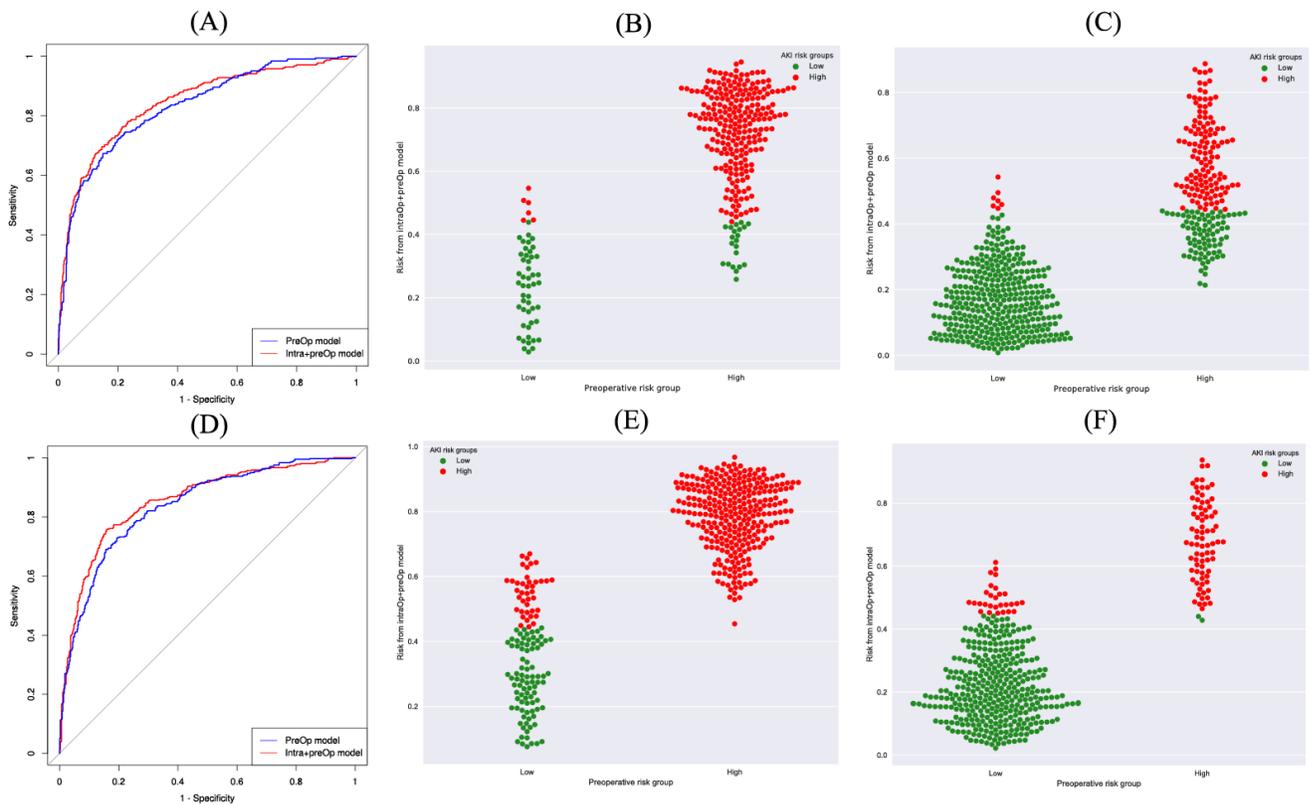

SDC Table 1. Characteristics of input variables.

| Variable | Type of Variable | Data Source | Number of categories | Type of Preprocessing |
|---|---|---|---|---|
| **Demographic variables** | | | | |
| Age (years) | Continuous | Derived | | Imputation of outliers[a]; Nonlinear function[b] |
| Gender | Binary | Raw | 2 | |
| Race | Nominal | Raw | 5 | Optimization of categorical features[c] |
| **Socioeconomic variables** | | | | |
| Primary Insurance | Nominal | Raw | 4 | Optimization of categorical features[c] |
| Residency area characteristics | | | | |
|     Zip code | Nominal | Raw | 10,000 | Transformation through link to Census data[d] |
|     County | Nominal | Raw | 71 | Optimization of categorical features[c] |
|     Rural area | Binary | Derived | 2 | |
|     Total Population | Continuous | Derived | | Obtained using residency zip code with linkage to US Census data[d]; Imputation of outliers[a] |
|     Median Income | Continuous | Derived | | Obtained using residency zip code with linkage to US Census data[d]; Imputation of outliers[a] |
|     Total Proportion of African-Americans | Continuous | Derived | | Obtained using residency zip code with linkage to US Census data[d]; Imputation of outliers[a] |
|     Total Proportion of Hispanic | Continuous | Derived | | Obtained using residency zip code with linkage to US Census data[d]; Imputation of outliers[a] |
|     Population Proportion Below Poverty | Continuous | Derived | | Obtained using residency zip code with linkage to US Census data[d]; Imputation of outliers[a] |
| Distance from Residency to Hospital (km) | Continuous | Derived | | Calculated using residency zip code; Imputation of outliers[a] |
| **Operative characteristics** | | | | |
| Day of admission | Nominal | Derived | 7 | Optimization of categorical features[c] |
| Month of admission | Nominal | Derived | 12 | Optimization of categorical features[c] |
| Year of admission | Nominal | Derived | 11 | Optimization of categorical features[c] |
| Weekend admission | Binary | Derived | 2 | |
| Attending Surgeon | Nominal | Raw | 520 | Optimization of categorical features[c] |
| Admission Source | Nominal | Raw | 3 | Optimization of categorical features[c] |
| Admission Type (Emergent/Elective) | Binary | Raw | 2 | |
| Admitting type (Medicine/Surgery) | Binary | Derived | 2 | |
| Admitting Service | Nominal | Derived | 46 | Optimization of categorical features[c] |
| Surgery Type | Nominal | Derived | 5 | Optimization of categorical features[c] |
| Time of surgery from admission (days) | Continuous | Derived | | Imputation of outliers[a]; Nonlinear function[b] |
| Diagnosis/Procedure | | | | |
|     Primary surgical procedure | Nominal | Derived | 1555 | Forest tree analysis of ICD9 codes[e] |
|     Major Diagnosis Category | Nominal | Raw | 28 | Optimization of categorical features[c] |
| **Comorbidities** | | | | |
| Charlson's comorbidity index | Nominal | Derived | 18 | Optimization of categorical features[c] |
| Number of diagnosis | Continuous | Derived | | Imputation of outliers[a]; Nonlinear function[b] |
| Myocardial Infarction | Binary | Derived | 2 | |
| Congestive Heart Failure | Binary | Derived | 2 | |
| Peripheral Vascular Disease | Binary | Derived | 2 | |
| Cerebrovascular Disease | Binary | Derived | 2 | |
| Chronic Pulmonary Disease | Binary | Derived | 2 | |
| Diabetes | Binary | Derived | 2 | |
| Cancer | Binary | Derived | 2 | |
| Liver Disease | Binary | Derived | 2 | |
| Valvular disease | Binary | Derived | 2 | |
| Hypothyroidism | Binary | Derived | 2 | |
| Coagulopthy | Binary | Derived | 2 | |
| Obesity | Binary | Derived | 2 | |
| Weight loss | Binary | Derived | 2 | |
| Fluid and electrolyte disorders | Binary | Derived | 2 | |
| Chronic anemia | Binary | Derived | 2 | |
| Alcohol or drug abuse | Binary | Derived | 2 | |
| Depression | Binary | Derived | 2 | |
| Hypertension | Binary | Derived | 2 | |
| Chronic kidney disease | Binary | Derived | 2 | |
| End Stage Renal disease | Binary | Derived | 2 | |
| **Admission day Medications[f]** | | | | |
| Number of Medications on Admission | Continuous | Derived | | Imputation of outliers[a]; Nonlinear function[b] |
| Betablockers | Binary | Derived | 2 | |
| Diuretics | Binary | Derived | 2 | |
| Statin | Binary | Derived | 2 | |
| Aspirin | Binary | Derived | 2 | |
| Angiotensin-Converting-Enzyme Inhibitors | Binary | Derived | 2 | |
| Pressors or inotropes | Binary | Derived | 2 | |
| Bicarbonate | Binary | Derived | 2 | |
| Antiemetic | Binary | Derived | 2 | |
| Aminoglycosides | Binary | Derived | 2 | |
| Steroids | Binary | Derived | 2 | |
| Vancomycin | Binary | Derived | 2 | |
| Nonsteroidal anti-inflammatory drug | Binary | Derived | 2 | |
| Number of Nephrotoxic Medications | Continuous | Derived | | Imputation of outliers[a]; Nonlinear function[b] |
| **Preoperative laboratory results** | | | | |
| Reference estimated glomerular filtration rate | Continuous | Derived | | Imputation of outliers[a]; Nonlinear function[b] |
| Ratio of reference creatinine to MDRD Cr | Continuous | Derived | | Imputation of outliers[a] |
| Hemoglobin, g/dl | Continuous | Raw | | Imputation of outliers[a]; Nonlinear function[b] |
| Automated urinalysis, urine protein, mg/dL | Nominal | Derived | 3 | Optimization of categorical features[c] |
| Automated urinalysis, urine hemoglobin, mg/dL | Nominal | Derived | 3 | Optimization of categorical features[c] |
| Automated urinalysis, urine glucose, mg/dL | Nominal | Derived | 3 | Optimization of categorical features[c] |
| No of complete blood count tests | Nominal | Derived | 3 | Optimization of categorical features[c] |
| **Physiologic intraoperative time series** | | | | |
| Mean arterial blood pressure (Invasive), mmHg | Continuous | Raw | | Data cleaning[g]; Imputation of outliers[h]; Statistical features extraction[i] |
| Systolic blood pressure (Invasive), mmHg | Continuous | Raw | | Data cleaning[g]; Imputation of outliers[h]; Statistical features extraction[i] |
| Diastolic blood pressure (Invasive), mmHg | Continuous | Raw | | Data cleaning[g]; Imputation of outliers[h]; Statistical features extraction[i] |
| Minimum alveolar concentration | Continuous | Raw | | Data cleaning[g]; Imputation of outliers[h]; Statistical features extraction[i] |
| Heart rate, bpm | Continuous | Raw | | Data cleaning[g]; Imputation of outliers[h]; Statistical features extraction[i] |
| Peripheral capillary oxygen saturation (SPO2) | Continuous | Raw | | Data cleaning[g]; Imputation of outliers[h] |
| **Laboratory results on surgery** | | | | |
| pH | Continuous | Raw | | Missing data imputation[j]; Statistical features extraction[k]; Outlier removal[a] |
| Lactic acid, mmol/L | Continuous | Raw | | Missing data imputation[j]; Statistical features extraction[k]; Outlier removal[a] |
| Hemoglobin, g/dL | Continuous | Raw | | Missing data imputation[j]; Statistical features extraction[k]; Outlier removal[a] |
| Hematocrit, vol% | Continuous | Raw | | Missing data imputation[j]; Statistical features extraction[k]; Outlier removal[a] |
| Partial pressure of oxygen (PO2)- Arterial, mmHg | Continuous | Raw | | Missing data imputation[j]; Statistical features |

| | | | |
|---|---|---|---|
| | | | extraction[k]; Outlier removal[a] |
| Fraction of inspired oxygen (FIO2) | Continuous | Raw | Missing data imputation[j]; Statistical features extraction[k]; Outlier removal[a] |
| PF-ratio | Continuous | Derived | Missing data imputation[l]; Outlier removal[a] |
| Mean corpuscular hemoglobin, pg/[red cell] | Continuous | Raw | Missing data imputation[j]; Statistical features extraction[k]; Outlier removal[a] |
| Mean corpuscular hemoglobin concentration, g/dL | Continuous | Raw | Missing data imputation[j]; Statistical features extraction[k]; Outlier removal[a] |
| Mean corpuscular volume, fl/[red cell]L | Continuous | Raw | Missing data imputation[j]; Statistical features extraction[k]; Outlier removal[a] |
| Red Blood Cells, million cells/mcL | Continuous | Raw | Missing data imputation[j]; Statistical features extraction[k]; Outlier removal[a] |
| Red Cell Distribution Width, % | Continuous | Raw | Missing data imputation[j]; Statistical features extraction[k]; Outlier removal[a] |
| White Blood Cells, 10*9/L | Continuous | Raw | Missing data imputation[j]; Statistical features extraction[k]; Outlier removal[a] |
| Mean Platelet Volume, fL | Continuous | Raw | Missing data imputation[j]; Statistical features extraction[k]; Outlier removal[a] |
| Platelets Count, 10*9/L | Continuous | Raw | Missing data imputation[j]; Statistical features extraction[k]; Outlier removal[a] |
| Carboxyhemoglobin-Arterial, % | Continuous | Raw | Missing data imputation[j]; Statistical features extraction[k]; Outlier removal[a] |

| | | | | |
|---|---|---|---|---|
| oxygen Content-Arterial, % | Continuous | Raw | | Missing data imputation[j]; Statistical features extraction[k]; Outlier removal[a] |
| oxygen saturation, % | Continuous | Raw | | Missing data imputation[j]; Statistical features extraction[k]; Outlier removal[a] |
| Bicarbonate-Arterial, mmol/L | Continuous | Raw | | Missing data imputation[j]; Statistical features extraction[k]; Outlier removal[a] |
| Methemoglobin, % | Continuous | Raw | | Missing data imputation[j]; Statistical features extraction[k]; Outlier removal[a] |
| Partial pressure of Carbon dioxide, mmHg | Continuous | Raw | | Missing data imputation[j]; Statistical features extraction[k]; Outlier removal[a] |
| **Medications on surgery** | | | | |
| Pressors | Binary | Derived | 2 | |
| Diuretic | Binary | Derived | 2 | |
| **Other characteristics** | | | | |
| Anesthesia type | Nominal | Raw | 2 | |
| Duration of surgery, min | Continuous | Derived | | Outlier removal[a] |
| Estimated blood loss, mL | Continuous | Raw | | Outlier removal[a] |
| Urine output, mL | Continuous | Raw | | Outlier removal[a] |
| Fluid, mL | Continuous | Raw | | Outlier removal[a] |
| Blood product, mL | Continuous | Raw | | Outlier removal[a] |

Abbreviation: MDRD, Modification of Diet in Renal Disease.

MDRD formula for creatinine was calculated using $(186/GFR) * (0.742*[if\ female])*(1.21*[if\ black])* age^{(-0.203)}]^{(1/1.154)}$ assuming GFR=75.

PF-ratio was calculated using PO2/FIO2.

Different set of variables were kept in final models (preoperative or intraoperative) from the input set provided in the table.

[a] For continuous variables, observations that fell in the top and bottom 1% of the distribution were considered as outliers and imputed by neighborhood values (i.e., above 99% are imputed randomly from

a uniform distribution defined over [95%, 99.5%] percentiles and below 1% are imputed randomly from another uniform distribution defined over [0.5%, 5%] percentiles.

[b] Nonlinear risk function was calculated for continuous functions entered to the preoperative models.

[c] For categorical variables with more than five levels, levels were transformed to a numeric value as detailed in Methods section.

[d] Using residency zip code, we linked to US Census data to calculate residing neighborhood characteristics and distance from hospital.

[e] Surgical procedure codes were optimized using forest tree analysis of ICD-9-CM codes as detailed in Methods section.

[f] Medications were dispensed on the first admission day using RxNorms data grouped into drug classes according to the US, Department of Veterans Affairs National Drug File-Reference Terminology.

[g] We used observations for the first surgery, in case multiple surgeries exist. We averaged values if multiple observations exist at a time point. Only accounts with more than 30 observations were considered.

[h] Values out of the predefined ranges were removed. Additionally, values in the top and bottom 0.5% of each of the time series distributions were removed.

[i] We extracted several descriptive statistical measures, i.e. mean and standard deviation of base signal and standard deviation of residual signal (Saria et al., 2010) of time series, minimum and maximum values observed, time/percentage of time a patient spent in a specific range of values for each of the time series.

[j] Observations during the first surgery have been considered. Missing values were imputed using surgery day observations for the given account.

[k] We extracted descriptive statistical features, i.e., minimum, mean, maximum, count, variance, abnormal observation percentage (%) = abnormal value count/ total observations * 100.

[l] Missing PF-ratios are replaced by (SF-ratio – 64)/0.84, where SF-ratio = SPO2/FIO2.

SDC Table 2. Summary of overall cohort stratified for AKI-3day and AKI-Overall by acute kidney injury.

| | Overall (N=2,911) | Acute Kidney Injury - 3day | | Acute Kidney Injury - Overall | |
|---|---|---|---|---|---|
| | | No (N=1,913) | Yes (N=998) | No (N=1,572) | Yes (N=1,339) |
| **Demographic features** | | | | | |
| Age, median (25th-75th) | 60.0 (49.0,69.0) | 58.0 (47.0,68.0) | 63.0 (52.0,72.0)[a] | 58.0 (46.0,67.0) | 63.0 (52.0,71.5)[a] |
| Male gender, n (%) | 1760 (60.4%) | 1101 (57.5%) | 659 (66.0%)[a] | 898 (57.1 %) | 862 (64.3 %)[a] |
| Race, n (%) | | | | | |
| White | 2374 (81.5%) | 1579 (82.5 %) | 795 (79.6 %) | 1295 (82.3 %) | 1079 (80.5 %) |
| AA | 265 (9.10%) | 167 (8.73 %) | 98 (9.82 %) | 133 (8.46 %) | 132 (9.86 %) |
| Hispanic | 117 (4.02%) | 66 (3.45 %) | 51 (5.11 %) | 57 (3.63 %) | 60 (4.48 %) |
| Missing | 87 (2.99%) | 52 (2.72 %) | 35 (3.51 %) | 44 (2.8 %) | 43 (3.21 %) |
| Other | 68 (2.34%) | 49 (2.56 %) | 19 (1.9 %) | 43 (2.74 %) | 25 (1.87 %) |
| Primary insurance, n (%)[a] | | | | | |
| Private | 1208 (41.5%) | 844 (44.1 %) | 364 (36.4 %) | 719 (45.7 %) | 489 (36.5 %) |
| Medicare | 1204 (41.3%) | 726 (37.9 %) | 478 (47.9 %) | 565 (35.9 %) | 639 (47.7 %) |
| Medicaid | 340 (11.68%) | 223 (11.66 %) | 117 (11.72 %) | 177 (11.26 %) | 163 (12.17 %) |
| Uninsured | 159 (5.46%) | 120 (6.27 %) | 39 (3.91 %) | 111 (7.06 %) | 48 (3.58 %) |
| Zip (top 3 codes), n (%) | | | | | |
| 32608 | 34 (1.17%) | 22 (1.15 %) | 12 (1.2 %) | 18 (1.15 %) | 16 (1.19 %) |
| 32060 | 27 (0.93%) | 16 (0.84 %) | 11 (1.1 %) | 13 (0.83 %) | 14 (1.05 %) |
| 32605 | 27 (0.93%) | 21 (1.1 %) | 6 (0.6 %) | 18 (1.15 %) | 9 (0.67 %) |
| **Socio-economic features** | | | | | |
| Neighborhood characteristics | | | | | |
| Rural area, n (%) | 767 (26.35%) | 503 (26.29 %) | 264 (26.45 %) | 431 (27.42 %) | 336 (25.09 %) |
| Total population, median (25th-75th) | 19162 (10639,30611) | 18931 (10510,30611) | 19363 (11056,30570) | 18931 (10448,30459) | 19363 (11116,30649) |
| Median income, median (25th-75th) | 34372 (29980,41410) | 34289 (29854,41410) | 34476 (30166,41191) | 34285 (29689,41363) | 34476 (30284,41410) |
| Total proportion of African-Americans (%), median (25th-75th) | 9.64 (3.91,17.6) | 9.54 (3.97,16.84) | 9.64 (3.80, 18.9) | 9.5 (3.91,16.6) | 9.64 (3.91,19.5) |
| Total proportion of Hispanic (%),median (25th-75th) | 4.27 (2.54,6.83) | 4.28 (2.55,6.73) | 4.07 (2.54,7.39) | 4.28( 2.65,6.72) | 4.19 (2.5,7.15) |
| County (top 3 categories), n(%) | | | | | |
| Alachua | 383 (13.16%) | 261 (13.64 %) | 122 (12.22 %)[a] | 221 (14.06 %) | 162 (12.1 %) |
| Marion | 250 (8.59%) | 190 (9.93 %) | 60 (6.01 %)[a] | 162 (10.31 %) | 88 (6.57 %) |
| Georgia | 142 (4.88%) | 88 (4.6 %) | 54 (5.41 %)[a] | 68 (4.33 %) | 74 (5.53 %) |
| Distance from residency to hospital (km), median (25th-75th) | 68.0 (28.9,143) | 61.7 (28.2,131) | 74.0 (30.7,156)[a] | 57.9 (28.2,130) | 72.1 (30.7,153)[a] |
| Population proportion below poverty (%), median (25th-75th) | 12.01 (8.18,17.36) | 12.02 (8.24,17.41) | 11.81 (8.04,17.23) | 12.11 (8.32,17.39) | 11.78 (8.04,17.27) |

*Preoperative characteristics of patients stratified by acute kidney injury*

**Comorbidity features**

| | | | | | |
|---|---|---|---|---|---|
| Charlson's comorbidity index (CCI), median (25th-75th) | 2.0 (1.0,3.0) | 1.0 (0.0,3.0) | 2.0 (1.0,3.0)[a] | 1.0 (0.0,3.0) | 2.0 (1.0,3.0)[a] |
| Chronic kidney disease, n (%) | 346 (11.89%) | 94 (4.91 %) | 252 (25.25 %)[a] | 52 (3.31 %) | 294 (21.96 %)[a] |

**Operative features**

**Admission**

| | | | | | |
|---|---|---|---|---|---|
| Weekend admission, n (%) | 472 (16.21%) | 281 (14.69 %) | 191 (19.14 %)[a] | 230 (14.63 %) | 242 (18.07 %)[a] |
| Admission source, n(%)[a] | | | | | |
|   Outpatient setting | 1753 (60.53%) | 1202 (63.3 %) | 551 (55.27 %) | 1017 (65.15 %) | 736 (55.13 %) |
|   Emergency room | 583 (20.13%) | 384 (20.22 %) | 199 (19.96 %) | 325 (20.82 %) | 258 (19.33 %) |
|   Transfer | 560 (19.34%) | 313 (16.48 %) | 247 (24.77 %) | 219 (14.03 %) | 341 (25.54 %) |
| Admission month (top 3 categories), n (%) | | | | | |
|   September | 279 (9.58%) | 184 (9.62 %) | 95 (9.52 %) | 149 (9.48 %) | 130 (9.71 %) |
|   October | 273 (9.38%) | 195 (10.19 %) | 78 (7.82 %) | 160 (10.18 %) | 113 (8.44 %) |
|   June | 254 (8.73%) | 178 (9.3 %) | 76 (7.62 %) | 148 (9.41 %) | 106 (7.92 %) |
| Number of operating surgeons, n | 129 | | | | |
| Number of procedures per operating surgeon, n (%)[a] | | | | | |
|   First rank | 422 (14.5%) | 211 (11.03 %) | 211 (21.14 %) | 162 (10.31 %) | 260 (19.42 %) |
|   Second rank | 267 (9.17%) | 207 (10.82 %) | 60 (6.01 %) | 178 (11.32 %) | 89 (6.65 %) |
|   Third rank | 258 (8.86%) | 144 (7.53 %) | 114 (11.42 %) | 113 (7.19 %) | 145 (10.83 %) |
| Admitting type, n (%)[a] | | | | | |
|   surgery | 2545 (87.43%) | 1725 (90.17 %) | 820 (82.16 %) | 1437 (91.41 %) | 1108 (82.75 %) |
|   medicine | 366 (12.57%) | 188 (9.83 %) | 178 (17.84 %) | 135 (8.59 %) | 231 (17.25 %) |
| Emergent surgery, n (%) | 1352 (46.44%) | 825 (43.13 %) | 527 (52.81 %)[a] | 644 (40.97 %) | 708 (52.88 %)[a] |
| Surgery type, n (%)[a] | | | | | |
|   Cardiothoracic surgery | 1415 (48.61%) | 782 (40.88 %) | 633 (63.43 %) | 596 (37.91 %) | 819 (61.17 %) |
|   Other surgeries | 826 (28.38%) | 569 (29.74 %) | 257 (25.75 %) | 477 (30.34 %) | 349 (26.06 %) |
|   Neurologic Surgery | 301 (10.34%) | 279 (14.58 %) | 22 (2.2 %) | 243 (15.46 %) | 58 (4.33 %) |
|   Specialty Surgeries | 243 (8.35%) | 193 (10.09 %) | 50 (5.01 %) | 181 (11.51 %) | 62 (4.63 %) |
|   Non-Cardiac General | 126 (4.33%) | 90 (4.7 %) | 36 (3.61 %) | 75 (4.77 %) | 51 (3.81 %) |
| Time of surgery from admission (days) | 0.0 (0.0,2.0) | 0.0 (0.0,1.0) | 1.0 (0.0,4.0)[a] | 0.0 (0.0,1.0) | 1.0 (0.0,4.0)[a] |

**Admission day medications**

| | | | | | |
|---|---|---|---|---|---|
| Diuretics | 600 (20.61%) | 323 (16.88 %) | 277 (27.76 %)[a] | 252 (16.03 %) | 348 (25.99 %)[a] |
| Bicarbonate | 295 (10.13%) | 162 (8.47 %) | 133 (13.33 %)[a] | 126 (8.02 %) | 169 (12.62 %)[a] |
| Angiotensin-Converting-Enzyme Inhibitors | 351 (12.06%) | 205 (10.72 %) | 146 (14.63 %)[a] | 163 (10.37 %) | 188 (14.04 %)[a] |
| Antiemetic | 1413 (48.54%) | 1008 (52.69 %) | 405 (40.58 %)[a] | 861 (54.77 %) | 552 (41.22 %)[a] |
| Betablockers | 872 (29.96%) | 561 (29.33 %) | 311 (31.16 %) | 460 (29.26 %) | 412 (30.77 %) |
| Statin | 502 (17.24%) | 298 (15.58 %) | 204 (20.44 %)[a] | 254 (16.16 %) | 248 (18.52 %) |
| Pressors or inotropes | 424 (14.57%) | 228 (11.92 %) | 196 (19.64 %)[a] | 173 (11.01 %) | 251 (18.75 %)[a] |

*Intraoperative characteristics of patients stratified by acute kidney injury*

**Physiologic intraoperative time series variables, median (25th-75th)**

| | | | | | |
|---|---|---|---|---|---|
| Diastolic BP - maximum | 140 (121,140) | 140 (120,140) | 140 (125,140)[a] | 140 (119,140) | 140 (125,140)[a] |
| Diastolic BP - minimum | 20.0 (15.0,25.0) | 20.0 (16.0,30.0) | 20.0 (13.0,24.0)[a] | 20.0 (16.8,30.0) | 20.0 (13.0,25)[a] |
| Diastolic BP - average | 58.7 (53.8,65.1) | 60.1 (55.2,66.6) | 56.5 (51.9,62.1)[a] | 60.9 (55.6,67.2) | 56.8 (52.4,62.3)[a] |
| Diastolic BP - long-term variability | 10.4 (8.7,12.7) | 10.3 (8.4,12.7) | 10.7 (9.0,12.7)[a] | 10.2 (8.4,12.6) | 10.7 (8.9,12.9)[a] |
| Diastolic BP - short-term variability | 4.8 (3.8,5.9) | 4.8 (3.7,5.9) | 5.0 (4.0,6.0)[a] | 4.7 (3.6,5.8) | 4.9 (4.0,5.95)[a] |
| Systolic BP - maximum | 240 (209,240) | 240 (205,240) | 240 (215,240) | 240 (204,240) | 240 (215,240)[a] |
| Systolic BP - minimum | 40.0 (31.0,50.0) | 40.0 (32.0,55.0) | 40.0 (29.3,41.0)[a] | 40.0 (33.0,58.0) | 40.0 (29.0,42.0)[a] |
| Systolic BP - average | 106 (93.8,118) | 110 (96.4,120) | 100 (90.3,113)[a] | 111 (97.7,121) | 101 (90.9,113)[a] |
| Systolic BP - long-term variability | 19.6 (15.6,24.2) | 18.8 (15.1,23.6) | 20.7 (16.9,25.0)[a] | 18.7 (15.0,23.3) | 20.5 (16.6,24.9)[a] |
| Systolic BP - short-term variability | 7.7 (6.1,9.5) | 7.7 (6.0,9.4) | 7.9 (6.4,9.6)[a] | 7.6 (5.8,9.4) | 7.9 (6.4,9.55)[a] |
| Mean Arterial BP - maximum | 240 (209,240) | 240 (205,240) | 240 (215,240) | 240 (204,240) | 240 (215,240)[a] |
| Mean Arterial BP - minimum | 40.0 (31.0,50.0) | 40.0 (32.0,55.0) | 40.0 (29.25,41)[a] | 40.0 (33.0,58.0) | 40.0 (29.0,42.0)[a] |
| Mean Arterial BP - average | 106 (93.8,118) | 110 (96.4,120) | 100 (90.3,113)[a] | 111 (97.7,121) | 101 (90.9,114)[a] |
| Mean Arterial BP - long-term variability | 19.6 (15.6,24.2) | 18.8 (15.1,23.6) | 20.7 (16.9,25.0)[a] | 18.7 (15.0,23.3) | 20.5 (16.6,24.9)[a] |
| Mean Arterial BP - short-term variability | 7.7 (6.1,9.5) | 7.7 (6.0,9.4) | 7.9 (6.4,9.6)[a] | 7.6 (5.8,9.4) | 7.9 (6.4,9.55)[a] |
| Heart rate - maximum | 146 (125,171) | 145 (124,168) | 148 (127,175)[a] | 143 (123,166) | 148 (129,175)[a] |
| Heart rate - minimum | 34.0 (18.0,53.0) | 38.0 (21.0,55.0) | 26.0 (17.0,48.0)[a] | 40.0 (22.0,55.0) | 27.0 (17.0,49.0)[a] |
| Heart rate - average | 82.8 (73.8,92.7) | 82.1 (73.5,92.0) | 84.0 (74.5,93.9)[a] | 81.5 (72.8,91.5) | 84.2 (75.1,94.1)[a] |
| Heart rate - long-term variability | 12.4 (8.6,19.3) | 11.6 (8.5,18.1) | 14.2 (9.22,21.6)[a] | 11.3 (8.4,17.2) | 14.0 (9.3,21.65)[a] |
| Heart rate - short-term variability | 5.5 (4.2,6.6) | 5.4 (4.0,6.5) | 5.75 (4.5,6.8)[a] | 5.3 (3.88,6.4) | 5.8 (4.5,6.8)[a] |
| MAC - maximum | 2.59 (1.71,3.42) | 2.56 (1.71,3.42) | 3.42 (1.91,3.42)[a] | 2.56 (1.71,3.42) | 3.42 (1.94,3.42)[a] |
| MAC - average | 0.6 (0.5,0.7) | 0.6 (0.5,0.7) | 0.6 (0.5,0.7) | 0.6 (0.5,0.7) | 0.6 (0.5,0.7) |
| MAC - long-term variability | 0.3 (0.2,0.4) | 0.3 (0.2,0.4) | 0.3 (0.2,0.5)[a] | 0.3 (0.2,0.4) | 0.3 (0.2,0.5)[a] |

**Intraoperative laboratory results, median (25th-75th)**

| | | | | | |
|---|---|---|---|---|---|
| Carboxyhemoglobin in arterial - maximum | 2.2 (1.4,3.0) | 2.1 (1.3,2.9) | 2.3 (1.5,3.2)[a] | 2.05 (1.3,2.9) | 2.3 (1.4,3.1)[a] |
| Carboxyhemoglobin in arterial - mean | 1.95 (1.3,2.75) | 1.86 (1.2,2.7) | 2.1 (1.4,2.9)[a] | 1.8 (1.2,2.66) | 2.1 (1.38,2.81)[a] |
| Carboxyhemoglobin in arterial - minimum | 1.6 (1.1,2.55) | 1.5 (1.0,2.5) | 1.8 (1.2,2.7)[a] | 1.5 (1.0,2.5) | 1.8 (1.1,2.6)[a] |
| Bicarbonate in Arterial - maximum | 23.3 (21.7,25.0) | 23.3 (21.7,24.9) | 23.3 (21.6,25.2) | 23.3 (21.7,24.9) | 23.3 (21.6,25.2) |
| Bicarbonate in Arterial - mean | 22.7 (21.2,24.4) | 22.7 (21.3,24.4) | 22.5 (21.0,24.4) | 22.7 (21.3,24.4) | 22.7 (21.0,24.4) |
| Bicarbonate in Arterial - mean | 22.0 (20.4,24.1) | 22.7 (21.3,24.4) | 22.5 (21.0,24.4) | 22.7 (21.3,24.4) | 22.7 (21.0,24.4) |
| Bicarbonate in Arterial - variance | 0.0 (0.0,0.84) | 0.0 (0.0,0.71) | 0.0 (0.0,1.34) | 0.0 (0.0,0.69) | 0.0 (0.0,1.22) |
| Hematocrit - maximum | 31.5 (28.7,34.9) | 31.5 (28.8,34.9) | 31.4 (28.5,34.9) | 31.7 (28.9,35.1) | 31.4 (28.5,34.7)[a] |
| Hematocrit - mean | 30.8 (28.0,34.1) | 30.9 (28.2,34.2) | 30.6 (27.7,34.0)[a] | 30.9 (28.2,34.3) | 30.6 (27.8,33.8)[a] |
| Hematocrit -minimum | 30.2 (27.1,33.6) | 30.4 (27.4,33.7) | 29.8 (26.6,33.4)[a] | 30.6 (27.5,33.9) | 29.9 (26.7,33.3)[a] |
| Hematocrit - variance | 0.0 (0.0,1.12) | 0.0 (0.0,0.98) | 0.0 (0.0,1.47) | 0.0 (0.0,1.12) | 0.0 (0.0,1.16) |
| Hemoglobin in arterial - maximum | 11.0 (10.0,12.2) | 11.1 (10.1,12.2) | 11.0 (9.93,12.2) | 11.1 (10.2,12.3) | 11.0 (9.9,12.2)[a] |
| Hemoglobin in arterial - mean | 10.55 (9.6,11.8) | 10.62 (9.7,11.85) | 10.49 (9.5,11.6)[a] | 10.7 (9.76,11.9) | 10.47 (9.5,11.6)[a] |
| Hemoglobin in arterial - mimimum | 10.2 (9.0,11.6) | 10.3 (9.1,11.6) | 10.1 (8.7,11.5)[a] | 10.4 (9.2,11.7) | 10.1 (8.8,11.4)[a] |
| Hemoglobin in arterial - variance | 0.0 (0.0,0.44) | 0.0 (0.0,0.39) | 0.0 (0.0,0.6) | 0.0 (0.0,0.36) | 0.0 (0.0,0.6) |
| Lactic acid - maximum | 2.5 (1.5,4.2) | 2.1 (1.4,3.4) | 3.55 (2.1,5.4)[a] | 2.0 (1.3,3.2) | 3.3 (1.9,5.15)[a] |
| Lactic acid - mean | 2.2 (1.4,3.57) | 1.8 (1.2,2.9) | 2.94 (1.88,4.7)[a] | 1.7 (1.2,2.77) | 2.8 (1.75,4.43)[a] |
| Lactic acid - minimum | 1.6 (1.1,2.9) | 1.4 (1.0,2.5) | 2.2 (1.3,4.2)[a] | 1.4 (1.0,2.32) | 2.1 (1.3,3.7)[a] |
| Lactic acid -variance | 0.0 (0.0,0.25) | 0.0 (0.0,0.18) | 0.0 (0.0,0.5)[a] | 0.0 (0.0,0.16) | 0.0 (0.0,0.5)[a] |
| Mean corpuscular hemoglobin concentration - maximum | 34.5 (33.5,35.5) | 34.4 (33.5,35.5) | 34.5 (33.6,35.5) | 34.4 (33.5,35.5) | 34.5 (33.5,35.5) |
| Mean corpuscular hemoglobin concentration - mean | 34.3 (33.4,35.3) | 34.3 (33.35,35.3) | 34.4 (33.5,35.3) | 34.3 (33.4,35.3) | 34.3 (33.5,35.3) |
| Mean corpuscular hemoglobin concentration - minimum | 34.1 (33.2,35.1) | 34.1 (33.2,35.1) | 34.2 (33.3,35.1) | 34.1 (33.2,35.1) | 34.2 (33.3,35.1) |
| Mean corpuscular hemoglobin concentration - variance | 0.0 (0.0,0.08) | 0.0 (0.0,0.08) | 0.0 (0.0,0.09) | 0.0 (0.0,0.08) | 0.0 (0.0,0.05)[a] |
| Mean corpuscular hemoglobin - maximum | 30.6 (29.4,31.8) | 30.6 (29.4,31.8) | 30.5 (29.4,31.8) | 30.6 (29.4,31.8) | 30.5 (29.4,31.8) |
| Mean corpuscular hemoglobin - mean | 30.4 (29.3,31.67) | 30.5 (29.2,31.7) | 30.4 (29.3,31.6) | 30.4 (29.3,31.7) | 30.4 (29.3,31.6) |
| Mean corpuscular hemoglobin - minimum | 30.3 (29.1,31.5) | 30.3 (29.1,31.6) | 30.2 (29.2,31.5) | 30.3 (29.1,31.5) | 30.3 (29.1,31.5) |
| Mean corpuscular hemoglobin - variance | 0.0 (0.0,0.04) | 0.0 (0.0,0.04) | 0.0 (0.0,0.05) | 0.0 (0.0,0.05) | 0.0 (0.0,0.04) |
| Mean corpuscular volume - maximum | 88.6 (85.2,92.2) | 88.4 (85.1,92.3) | 88.7 (85.4,91.98) | 88.4 (85.2,92.2) | 88.7 (85.3,92.2) |
| Mean corpuscular volume - mean | 88.2 (85.0,91.8) | 88.2 (85.0,92.0) | 88.3 (85.2,91.4) | 88.1 (85.1,91.9) | 88.3 (85.0,91.7) |
| Mean corpuscular volume - minimum | 87.9 (84.7,91.4) | 87.9 (84.7,91.8) | 87.9 (84.8,90.9) | 87.9 (84.9,91.6) | 88.0 (84.7,91.3) |
| Mean corpuscular volume - variance | 0.0 (0.0,0.18) | 0.0 (0.0,0.18) | 0.0 (0.0,0.25) | 0.0 (0.0,0.2) | 0.0 (0.0,0.18) |
| Methemoglobin - max | 0.8 (0.6,1.0) | 0.8 (0.6,0.9) | 0.8 (0.6,1.0)[a] | 0.8 (0.6,0.9) | 0.8 (0.6,1.0)[a] |
| Methemoglobin - mean | 0.7 (0.52,0.9) | 0.7 (0.5,0.83) | 0.7 (0.55,0.9)[a] | 0.7 (0.5,0.8) | 0.7 (0.56,0.9)[a] |
| Methemoglobin - minimum | 0.6 (0.4,0.8) | 0.6 (0.4,0.8) | 0.6 (0.4,0.9)[a] | 0.6 (0.4,0.8) | 0.6 (0.4,0.9)[a] |
| Methemoglobin - variance | 0.0 (0.0,0.02) | 0.0 (0.0,0.02) | 0.0 (0.0,0.03) | 0.0 (0.0,0.02) | 0.0 (0.0,0.03) |
| Mean Platelet Volume - maximum | 8.0 (7.5,8.7) | 7.9 (7.4,8.5) | 8.2 (7.6,8.9)[a] | 7.9 (7.4,8.5) | 8.2 (7.6,8.8)[a] |
| Mean Platelet Volume - mean | 7.9 (7.4,8.5) | 7.8 (7.35,8.4) | 8.05 (7.5,8.7)[a] | 7.8 (7.3,8.35) | 8.03 (7.5,8.7)[a] |
| Mean Platelet Volume - minimum | 7.8 (7.3,8.4) | 7.7 (7.2,8.3) | 7.9 (7.3,8.57)[a] | 7.7 (7.2,8.2) | 7.9 (7.4,8.6)[a] |
| Mean Platelet Volume - variance | 0.0 (0.0,0.03) | 0.0 (0.0,0.03) | 0.0 (0.0,0.02) | 0.0 (0.0,0.04) | 0.0 (0.0,0.02) |
| O2 Content, Arterial - maximum | 15.3 (14.0,16.9) | 15.3 (14.0,17.0) | 15.2 (13.9,16.9) | 15.4 (14.1,17.0) | 15.2 (13.9,16.9)[a] |
| O2 Content, Arterial - mean | 14.7 (13.4,16.3) | 14.8 (13.5,16.4) | 14.5 (13.2,16.1)[a] | 14.8 (13.5,16.5) | 14.5 (13.2,16.1)[a] |
| O2 Content, Arterial - minimum | 14.2 (12.5,16.0) | 14.3 (12.8,16.1) | 13.9 (12.2,15.8)[a] | 14.4 (12.9,16.1) | 14.0 (12.3,15.8)[a] |
| O2 Content, Arterial - variance | 0.0 (0.0,0.91) | 0.0 (0.0,0.76) | 0.0 (0.0,1.23) | 0.0 (0.0,0.74) | 0.0 (0.0,1.11) |
| O2 saturation - max | 99.8 (99.0,100) | 99.8 (99.0,100) | 99.9 (99.0,100) | 99.8 (98.9,100) | 99.9 (99.0,100)[a] |
| O2 saturation - mean | 99.5 (98.4,100) | 99.45 (98.4,100) | 99.6 (98.4,100) | 99.4 (98.3,100) | 99.6 (98.5,100)[a] |
| O2 saturation - minimum | 99.3 (97.8,100) | 99.3 (97.8,100) | 99.4 (97.7,100) | 99.2 (97.7,100) | 99.4 (97.8,100)[a] |
| O2 saturation - variance | 0.0 (0.0,0.02) | 0.0 (0.0,0.02) | 0.0 (0.0,0.01) | 0.0 (0.0,0.02) | 0.0 (0.0,0.01) |
| Partial pressure of carbon dioxide - maximum | 42.8 (38.4,47.2) | 42.8 (38.2,47.0) | 43.4 (38.7,47.6)[a] | 42.8 (38.2,46.9) | 43.1 (38.7,47.5)[a] |
| Partial pressure of carbon dioxide - mean | 40.8 (36.9,45.2) | 40.8 (36.8,45.2) | 40.9 (37.1,45.2) | 40.8 (36.7,45.0) | 40.8 (37.2,45.4) |
| Partial pressure of carbon dioxide - minimum | 38.9 (34.55,44.6) | 38.8 (34.5,44.5) | 39.3 (34.7,44.77) | 38.8 (34.3,44.4) | 39.2 (34.7,44.8) |
| Partial pressure of carbon dioxide - variance | 0.0 (0.0,6.5) | 0.0 (0.0,5.89) | 0.0 (0.0,7.2) | 0.0 (0.0,5.91) | 0.0 (0.0,6.87) |
| PF-ratio | 90.4 (74.9,118) | 91.7 (75.4,122) | 87.6 (74.4,113)[a] | 91.4 (75.4,122) | 89.3 (74.4,115)[a] |
| pH - maximum | 7.38 (7.33,7.43) | 7.38 (7.34,7.43) | 7.38 (7.32,7.43) | 7.38 (7.34,7.43) | 7.38 (7.33,7.43) |
| pH - mean | 7.36 (7.32,7.4) | 7.36 (7.33,7.41) | 7.36 (7.31,7.4)[a] | 7.36 (7.33,7.41) | 7.36 (7.32,7.4)[a] |
| pH - minimum | 7.34 (7.3,7.39) | 7.34 (7.31,7.39) | 7.34 (7.29,7.38)[a] | 7.35 (7.31,7.39) | 7.34 (7.29,7.38)[a] |
| Platelet count - maximum | 180 (133,239) | 194 (143,251) | 161 (116,207)[a] | 198 (147,254) | 165 (118,211)[a] |
| Platelet count - mean | 176 (128,229) | 189 (139,243) | 154 (109,201)[a] | 192 (142,246) | 159 (112,207)[a] |
| Platelet count - minimum | 171 (121,223) | 183 (133,238) | 149 (102,194)[a] | 185 (137,242) | 155 (106,202)[a] |
| Platelet count - variance | 0.0 (0.0,78.1) | 0.0 (0.0,74.3) | 0.0 (0.0,82.7) | 0.0 (0.0,98.0) | 0.0 (0.0,60.5)[a] |
| Red blood cells - maximum | 3.59 (3.23,3.97) | 3.6 (3.25,3.98) | 3.58 (3.22,3.96) | 3.6 (3.26,3.99) | 3.57 (3.21,3.94)[a] |
| Red blood cells - mean | 3.5 (3.16,3.87) | 3.51 (3.18,3.88) | 3.48 (3.11,3.86)[a] | 3.53 (3.19,3.9) | 3.46 (3.12,3.85)[a] |
| Red blood cells - minimum | 3.42 (3.07,3.83) | 3.44 (3.1,3.84) | 3.38 (3.0,3.81)[a] | 3.46 (3.12,3.86) | 3.38 (3.01,3.8)[a] |
| Red blood cells - variance | 0.0 (0.0,0.02) | 0.0 (0.0,0.01) | 0.0 (0.0,0.02) | 0.0 (0.0,0.01) | 0.0 (0.0,0.02) |
| Red cell distribution width - maximum | 14.9 (13.9,16.1) | 14.6 (13.7,15.8) | 15.3 (14.4,16.7)[a] | 14.5 (13.6,15.6) | 15.3 (14.4,16.6)[a] |
| Red cell distribution width - mean | 14.8 (13.8,15.9) | 14.6 (13.7,15.7) | 15.2 (14.4,16.5)[a] | 14.4 (13.6,15.5) | 15.2 (14.3,16.5)[a] |
| Red cell distribution width - minimum | 14.7 (13.8,15.9) | 14.5 (13.6,15.6) | 15.1 (14.2,16.3)[a] | 14.3 (13.6,15.3) | 15.1 (14.2,16.3)[a] |
| Red cell distribution width - variance | 0.0 (0.0,0.01) | 0.0 (0.0,0.01) | 0.0 (0.0,0.02) | 0.0 (0.0,0.01) | 0.0 (0.0,0.01) |
| White blood cells - maximum | 13.6 (9.85,18.3) | 13.3 (9.9,17.9) | 14.05 (9.7,19.1)[a] | 13.2 (9.8,17.7) | 13.9 (9.9,19.1)[a] |
| White blood cells - mean | 12.9 (9.3,17.5) | 12.7 (9.3,17.1) | 13.3 (9.12,17.9) | 12.4 (9.3,16.8) | 13.3 (9.3,18.3)[a] |
| White blood cells - min | 12.1 (8.5,16.9) | 11.8 (8.6,16.6) | 12.6 (8.3,17.5) | 11.6 (8.5,16.3) | 12.7 (8.4,17.6)[a] |
| White blood cells - variance | 0.0 (0.0,0.85) | 0.0 (0.0,0.86) | 0.0 (0.0,0.85) | 0.0 (0.0,1.12) | 0.0 (0.0,0.54)[a] |
| Total blood products in ml | 318 (0.0,1250) | 0.0 (0.0,750) | 793 (0.0,2750)[a] | 0.0 (0.0,727) | 750 (0.0,2486)[a] |
| Total estimated blood loss in ml | 150 (150,500) | 150 (150,500) | 150 (150,388)[a] | 150 (150,500) | 150 (150,400)[a] |
| Total fluids in ml | 2300 (1400,3600) | 2400 (1500,3600) | 2200 (1250,3500) | 2400 (1500,3550) | 2200 (1250,3600) |
| Total urine output in ml | 650 (300,1150) | 650 (300,1100) | 650 (346,1200) | 600 (300,1100) | 700 (350,1200)[a] |
| **Intraoperative medications, n (%)** | | | | | |
| Diuretic (vs. No) | 329 (11.3%) | 141 (7.37 %) | 188 (18.84 %)[a] | 100 (6.36 %) | 229 (17.1 %)[a] |
| Pressors (vs. No) | 1942 (66.71%) | 1201 (62.8 %) | 741 (74.3 %)[a] | 970 (61.7 %) | 972 (72.6 %)[a] |
| **Other variables** | | | | | |
| Total surgery time in mins, median (25th-75th) | 387 (294,483) | 366 (275,455) | 427 (338,525)[a] | 352 (265,443) | 424 (338,52)[a] |
| Night surgery (vs. No), n (%) | 394 (13.53%) | 238 (12.44 %) | 156 (15.63 %)[a] | 167 (10.62 %) | 227 (16.95 %)[a] |
| General anesthesia (vs. BL/BM), n (%) | 2881 (98.97%) | 1893 (98.95 %) | 988 (99.0 %) | 1553 (98.79 %) | 1328 (99.18 %) |

[a] p-value < 0.05 compared to no AKI group.

# REFERENCES

1. Hobson C, Singhania G, Bihorac A. Acute Kidney Injury in the Surgical Patient. Crit Care Clin. 2015;31(4):705-+.
2. Chertow GM, Burdick E, Honour M, Bonventre JV, Bates DW. Acute kidney injury, mortality, length of stay, and costs in hospitalized patients. Journal of the American Society of Nephrology. 2005;16(11):3365-70.
3. Hobson CE, Yavas S, Segal MS, Schold JD, Tribble CG, Layon AJ, et al. Acute Kidney Injury Is Associated With Increased Long-Term Mortality After Cardiothoracic Surgery. Circulation. 2009;119(18):2444-53.
4. Brown JR, Rezaee ME, Marshall EJ, Matheny ME. Hospital Mortality in the United States following Acute Kidney Injury. Biomed Res Int. 2016;2016:4278579.
5. de Geus HR, Betjes MG, Bakker J. Biomarkers for the prediction of acute kidney injury: a narrative review on current status and future challenges. Clin Kidney J. 2012;5(2):102-8.
6. Ostermann M, Joannidis M. Biomarkers for AKI improve clinical practice: no. Intensive Care Medicine. 2015;41(4):618-22.
7. Koyner JL, Adhikari R, Edelson DP, Churpek MM. Development of a Multicenter Ward-Based AKI Prediction Model. Clin J Am Soc Nephro. 2016;11(11):1935-43.
8. Bihorac A, Brennan M, Ozrazgat-Baslanti T, Bozorgmehri S, Efron PA, Moore FA, et al. National surgical quality improvement program underestimates the risk associated with mild and moderate postoperative acute kidney injury. Crit Care Med. 2013;41(11):2570-83.
9. Birnie K, Verheyden V, Pagano D, Bhabra M, Tilling K, Sterne JA, et al. Predictive models for kidney disease: improving global outcomes (KDIGO) defined acute kidney injury in UK cardiac surgery. Critical Care. 2014;18(6).
10. Ng SY, Sanagou M, Wolfe R, Cochrane A, Smith JA, Reid CM. Prediction of acute kidney injury within 30 days of cardiac surgery. Journal of Thoracic and Cardiovascular Surgery. 2014;147(6):1875-U225.
11. Flechet M, Guiza F, Schetz M, Wouters P, Vanhorebeek I, Derese I, et al. AKIpredictor, an online prognostic calculator for acute kidney injury in adult critically ill patients: development, validation and comparison to serum neutrophil gelatinase-associated lipocalin. Intensive Care Med. 2017;43(6):764-73.
12. Walsh M, Devereaux PJ, Garg AX, Kurz A, Turan A, Rodseth RN, et al. Relationship between Intraoperative Mean Arterial Pressure and Clinical Outcomes after Noncardiac Surgery: Toward an Empirical Definition of Hypotension. Anesthesiology. 2013;119(3):507-15.
13. Haase M, Bellomo R, Story D, Letis A, Klemz K, Matalanis G, et al. Effect of mean arterial pressure, haemoglobin and blood transfusion during cardiopulmonary bypass on post-operative acute kidney injury. Nephrology Dialysis Transplantation. 2012;27(1):153-60.
14. Kashani K, Steuernagle JH, Akhoundi A, Alsara A, Hanson AC, Kor DJ. Vascular Surgery Kidney Injury Predictive Score: A Historical Cohort Study. Journal of Cardiothoracic and Vascular Anesthesia. 2015;29(6):1588-95.
15. Saria S, Rajani AK, Gould J, Koller D, Penn AA. Integration of early physiological responses predicts later illness severity in preterm infants. Sci Transl Med. 2010;2(48):48ra65.

16. Bihorac A, Ozrazgat-Baslanti T, Ebadi A, Motaei A, Madkour M, Pardalos PM, et al. MySurgeryRisk: Development and Validation of a Machine-Learning Risk Algorithm for Major Complications and Death after Surgery. Annals of Surgery. 2018.
17. Hastie T, Tibshirani, R, Friedman, JH. The elements of statistical learning: data mining, inference, and prediction. 2nd ed: Springer, New York, NY; 2009.
18. Hastie T, Tibshirani, R. Generalized Additive Models. 1st ed: Chapman and Hall; 1990.
19. Wood SN. Thin plate regression splines. Journal of the Royal Statistical Society Series B-Statistical Methodology. 2003;65:95-114.
20. Breiman L. Random Forests. Machine Learning. 2001;45(1):5-32.
21. Pedregosa F, Varoquaux G, Gramfort A, Michel V, Thirion B, Grisel O, et al. Scikit-learn: Machine Learning in Python. Journal of Machine Learning Research. 2011;12:2825-30.
22. Youden WJ. Index for Rating Diagnostic Tests. Cancer. 1950;3(1):32-5.
23. Pencina MJ, D'Agostino RB, Sr., Steyerberg EW. Extensions of net reclassification improvement calculations to measure usefulness of new biomarkers. Stat Med. 2011;30(1):11-21.